\documentclass[journal]{IEEEtran}

\usepackage{amsthm}
\newtheorem{theorem}{Theorem}

\newtheorem{lemma}{Lemma}
\newtheorem{definition}{Definition}

\newtheorem{assumption}{Assumption}
\usepackage{algorithm}
\usepackage{algpseudocode}
\usepackage{amsmath,amsfonts,amssymb}
\allowdisplaybreaks[4]
\usepackage{array}
\usepackage{cite}
\usepackage[caption=false,font=normalsize,labelfont=sf,textfont=sf]{subfig}
\usepackage{textcomp}
\usepackage{stfloats}
\usepackage{url}
\usepackage{verbatim}
\usepackage{graphicx}
\usepackage{multirow}
\usepackage{nomencl}
\makenomenclature
\usepackage{etoolbox}
\usepackage{booktabs}
\usepackage{xcolor}
\usepackage{tikz}
\usepackage{tabularx}

\newcommand{\rev}[1]{{#1}}

\begin{document}

\title{A Mixed $\mathcal{H}_{\infty}$--Passivity Approach for Leveraging
Heat-Pump-Connected District Heating Systems as Frequency Support in Electric Power Systems}
\author{Xinyi Yi,~\IEEEmembership{Student Member,~IEEE,}
        Ioannis Lestas,~\IEEEmembership{Member,~IEEE}
\thanks{X. Yi and I. Lestas are with the Department of Engineering, University of Cambridge, Trumpington Street, Cambridge, CB2 1PZ, United Kingdom. Emails:
        {\tt\small <xy343, icl20>@cam.ac.uk}. Conference paper \cite{yi2026mixedhinftypassivityapproachleveraging} is a preliminary version.
This manuscript extends it by incorporating
heat-disturbance and DHS-regulation channels into the multi-channel
$\mathcal H_\infty$ attenuation design, enabling more comprehensive analysis of dynamic interactions
and disturbance propagation within the EPS--HP--DHS system,
together with cross-model validation under more detailed electro--thermal dynamics.}
        }

\maketitle

\begin{abstract}
Heat-pump (HP)-connected district heating systems (DHSs) can support grid
frequency by rapidly modulating HP electrical consumption while accommodating
the resulting heat imbalance. Thermal-state-informed HP control improves
EPS--DHS coordination, but introduces slower DHS dynamics and heat-side
disturbance propagation at the electrical interface on the time scale of
secondary frequency control. This paper formulates the regulated DHS as a
grid-facing dynamic port and develops a mixed $\mathcal H_\infty$--passivity
framework for disturbance-independent temperature regulation while preserving
a clean electrical response during secondary frequency control. Sufficient
conditions guarantee local asymptotic stability of the EPS--HP--DHS
interconnection across primary frequency control, secondary frequency control,
and DHS regulation. These stability conditions are integrated with
multi-channel $\mathcal H_\infty$ attenuation of DHS-state-dependent electrical
dynamics, heat-disturbance propagation, and DHS-regulation errors, yielding a
tractable convex LMI program for controller synthesis. Cross-model validation deploys controllers synthesized on control-oriented
EPS--DHS models, without retuning, on an open-source electromechanical IEEE
39-bus model coupled with four high-fidelity finite-volume DHS models retaining
thermal transport, storage, and heat-loss dynamics. The resulting controller improves
EPS--DHS coordination without compromising the grid-facing electrical response.
\end{abstract}

\begin{IEEEkeywords}
Frequency support, district heating systems, heat pumps, thermal-state-informed
control, passivity-based control, $\mathcal H_\infty$ loop shaping,
power--heat systems.
\end{IEEEkeywords}

\IEEEpeerreviewmaketitle

\section{Introduction}
Rapid growth in electricity demand from data centers and electrified heating is increasing the need for flexible resources for power-system frequency support. Heat pumps (HPs) connected to district heating systems (DHSs) can provide such flexibility by rapidly modulating electrical demand while leveraging DHS thermal inertia to absorb the resulting heat imbalances. Large-scale HP deployment in Europe and China~\cite{heatpumpaccel,chinaheatpump} further motivates their use for frequency support.

Conventional frequency-responsive HPs modulate electrical demand according to
grid frequency, while DHS thermal inertia accommodates the resulting heat
imbalance. Such direct frequency-responsive HP control provides a clean
grid-facing response, whereas incorporating DHS thermal states enables
coordinated electro--thermal operation and power sharing~\cite{qin2025_tpwrs_coupled}. This thermal-state-informed coupling, however, exposes slower DHS transport,
storage, and regulator dynamics, together with heat-side disturbances, at the
electrical interface on the time scale of secondary frequency control.

\begin{table}[htbp]
\centering
\caption{Comparison of representative HP frequency-support formulations,
highlighting the use of thermal-state feedback, grid-facing dynamic shaping,
and coupled EPS--DHS stability and regulation analysis.}
\label{tab:literature_comparison}

\scriptsize
\setlength{\tabcolsep}{1.5pt}
\renewcommand{\arraystretch}{1.06}

\begin{tabularx}{\columnwidth}{
p{0.16\columnwidth}
p{0.25\columnwidth}
>{\raggedright\arraybackslash}X
p{0.31\columnwidth}
}
\hline

Reference
&
HP control model
&
Dynamic shaping
&
Coupled-system analysis
\\
\hline

% =========================================================
% Direct frequency-responsive HP control
% =========================================================

\multicolumn{4}{l}{
\textit{Direct frequency-responsive HP control}
}
\\
\hline

\cite{kim2014modeling,rasmussen2021primary,
9703676,muhssin2018potential,
rodriguez2019heuristic,koch2026frequency}
&
Grid frequency
&
No explicit grid-facing shaping
&
No coupled EPS--DHS stability result
\\

\cite{qin2025_tpwrs_coupled}, Mode 1
&
Grid frequency
&
No explicit grid-facing shaping
&
Coupled stability under primary frequency control; optimal power sharing
\\

\hline

% =========================================================
% Thermal-state-informed HP control
% =========================================================

\multicolumn{4}{l}{
\textit{Thermal-state-informed HP control}
}
\\
\hline

\cite{muhssin2018dynamic}
&
Grid frequency + building temperature
&
No explicit grid-facing shaping
&
No coupled EPS--DHS stability result
\\

\cite{bunning2022robust}
&
Grid frequency + thermal-storage state
&
No explicit grid-facing shaping
&
No coupled EPS--DHS stability result
\\

\cite{yan2025demand}
&
Grid frequency + DHS supply-water temperature
&
No explicit grid-facing shaping
&
No coupled EPS--DHS stability result
\\

\cite{qin2025_tpwrs_coupled}, Mode 2
&
Grid frequency + DHS average temperature
&
No explicit grid-facing shaping
&
Coupled stability under primary frequency control;
electro--thermal power sharing
\\

\hline

% =========================================================
% HP actuation and internal dynamics
% =========================================================

\multicolumn{4}{l}{
\textit{HP actuation and internal dynamics}
}
\\
\hline

\cite{meesenburg2020optimizing,song2024integrating}
&
Grid command + internal HP states
&
Device-level electrical-response shaping
&
No coupled EPS--DHS stability result
\\

\hline

% =========================================================
% Coupled electro--thermal formulation
% =========================================================

\multicolumn{4}{l}{
\textit{Coupled electro--thermal system analysis}
}
\\
\hline

\cite{krishna2021port}
&
Electrical + thermal states
&
No explicit grid-facing DHS-port shaping
&
Stability under secondary frequency and temperature
regulation
\\

\hline

% =========================================================
% This work
% =========================================================

\textbf{This work}
&
\textbf{Grid frequency + regulated DHS states}
&
\textbf{Grid-facing regulated-DHS port shaping}
&
\textbf{EPS--HP--DHS stability under primary and secondary control;
DHS regulation and optimal heat sharing}
\\

\hline

\end{tabularx}
\end{table}

Existing HP frequency-support studies span layers of HP control and
dynamics, as summarized in Table~\ref{tab:literature_comparison}.
At the HP-control level, direct frequency-responsive HP control has been
studied for individual and aggregated HPs
~\cite{kim2014modeling,rasmussen2021primary,9703676,
muhssin2018potential,rodriguez2019heuristic,koch2026frequency}, while
thermal-state-informed HP control has used feedback signals based on building
temperature~\cite{muhssin2018dynamic}, thermal-storage states
~\cite{bunning2022robust}, and DHS supply-water temperature
~\cite{yan2025demand}. A separate line of work addresses HP actuation and
internal dynamics to enable fast electrical-power modulation
~\cite{meesenburg2020optimizing,song2024integrating}. At the coupled-system
level,~\cite{krishna2021port} developed a port-Hamiltonian model of the
electro--thermal system and established closed-loop stability under secondary
frequency and temperature control, while~\cite{qin2025_tpwrs_coupled}
established stability and power-sharing properties for primary frequency
support under both direct frequency-responsive and thermal-state-informed HP
coupling.

\begin{figure*}[t]
    \centering
    \includegraphics[width=\textwidth]{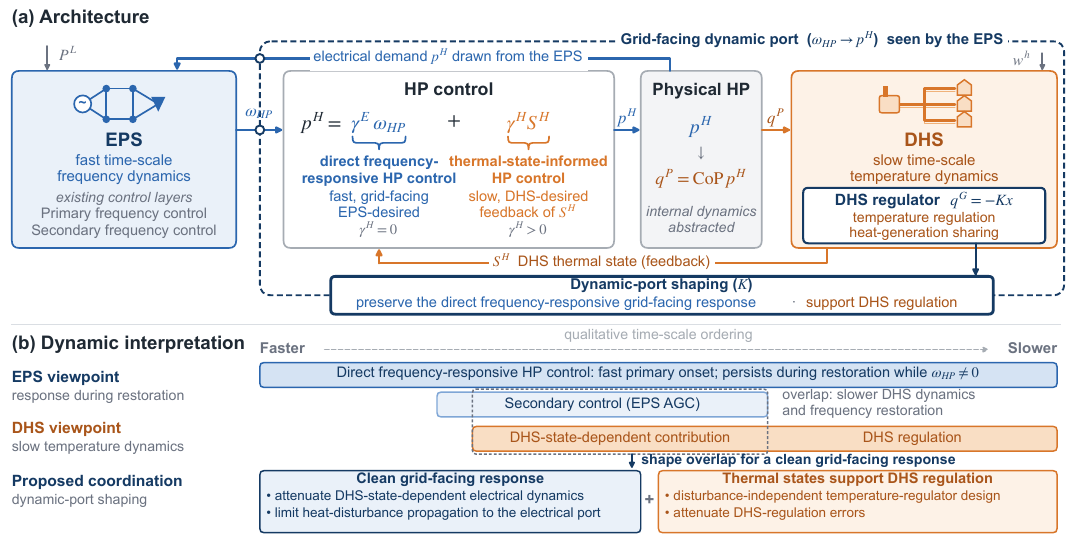}
    \caption{Overview of the EPS--HP--DHS architecture and its interpretation.
(a) Direct frequency-responsive HP control supplies the desired electrical
response; thermal-state-informed HP control supports DHS temperature regulation
and heat-generation sharing. Dynamic-port shaping limits the electrical impact
of DHS-state-dependent dynamics and heat-side disturbances.
(b) The direct HP response begins on the fast primary-control time scale and
persists while frequency deviation remains nonzero during secondary frequency
control. EPS AGC provides secondary control. Shaping targets the overlap of
slower DHS dynamics with secondary frequency control to keep the grid-facing
response clean. Bar extents are qualitative.}
    \label{fig:overview_pipeline}
\end{figure*}

The key challenge is to retain the DHS-regulation benefit of thermal-state-informed
HP control~\cite{qin2025_tpwrs_coupled} while preserving the desired electrical
behavior of direct frequency-responsive HP control. We model the regulated DHS
as a grid-facing dynamic port and shape this pathway to limit the electrical
impact of DHS-state-dependent dynamics and heat-side disturbances during
secondary frequency control. As illustrated in Fig.~\ref{fig:overview_pipeline},
the grid-facing response of the HP-connected DHS under thermal-state-informed HP control spans the 
primary-control and secondary-control time scales. The DHS-state-dependent
contribution overlaps with secondary frequency control, motivating shaping to
keep the grid-facing response clean. The coupled stability analysis spans
primary frequency control, secondary frequency control, and DHS regulation.

The DHS regulator acts within the DHS and shapes its thermal dynamics, which
are exposed to the EPS through thermal-state-informed HP control. Conventional
approaches~\cite{machado2022decentralized} compute optimal setpoints from
predicted heat disturbances and regulate the DHS around them, so disturbance
mismatch can shift the desired operating point. Our previous
work~\cite{yi2025optimal,yi2025policy} instead embeds temperature-regulation
and heat-generation-sharing optimality conditions into the closed-loop dynamics,
enabling convergence to the optimal equilibrium without heat-disturbance
prediction. This disturbance-independent regulator design provides the
DHS-regulation layer on which the present grid-facing dynamic-port shaping is
built.

Our preliminary work~\cite{yi2026mixedhinftypassivityapproachleveraging}
developed a mixed $\mathcal H_\infty$--passivity framework for DHS-regulator
synthesis by formulating the regulated DHS as a grid-facing dynamic port.
This paper extends that framework by incorporating heat-disturbance and
DHS-regulation channels into the multi-channel $\mathcal H_\infty$ attenuation
design, providing a more comprehensive analysis of dynamic interactions and
disturbance propagation within the interconnected EPS--HP--DHS system, together
with cross-model validation incorporating more detailed and realistic
electro--thermal dynamics.

The main contributions are as follows.
\begin{itemize}
\item \textbf{Grid-facing dynamic-port formulation:}
the regulated DHS under thermal-state-informed HP control is formulated as a
grid-facing dynamic port, enabling systematic analysis of the dynamic
interaction between DHS regulation and the HP grid-facing electrical response.

\item \textbf{Stability-certified multi-channel dynamic-port shaping:}
sufficient conditions are derived for local asymptotic stability of the
EPS--HP--DHS interconnection spanning the time scales of primary frequency
control, secondary frequency control, and DHS regulation. These conditions are
formulated as tractable LMIs and combined with a multi-channel
$\mathcal H_\infty$ design that attenuates DHS-state-dependent electrical
dynamics, heat-disturbance propagation to the electrical port, and
DHS-regulation errors.

\item \textbf{Cross-model validation under electro--thermal dynamics:}
controllers synthesized on control-oriented EPS--DHS models are deployed
without retuning on an electromechanical IEEE 39-bus model coupled with four
finite-volume DHS models~\cite{10380765} retaining thermal
transport, storage, and heat-loss dynamics. The results show that the controller
preserves the grid-facing response of direct frequency-responsive HP control
during secondary frequency control, and improves DHS regulation.
\end{itemize}

Section~\ref{sec:model} develops the EPS--HP--DHS model and
grid-facing dynamic-port formulation. Section~\ref{sec:nominal_synthesis}
establishes the equilibrium and stability conditions for the coupled system.
Section~\ref{sec:loop_shaping} develops the multi-channel
$\mathcal H_\infty$ attenuation design.
Section~\ref{sec:numerical} provides cross-model validation under
electro--thermal dynamics, and Section~\ref{sec:conclusion} concludes the paper.

\section{Coupled EPS--HP--DHS Model and Port Formulation}
\label{sec:model}

This section develops the coupled EPS--HP--DHS model for
port-based interconnection analysis. Unless stated otherwise, all variables denote deviations from a
prescribed operating point. Bold symbols denote vectors or matrices, whereas
non-bold symbols denote scalars.
\rev{Table~\ref{tab:notation} summarizes selected notation used throughout the
paper.}
For $\boldsymbol x\in\mathbb R^n$, $\|\boldsymbol x\|$ denotes the Euclidean
norm. For a continuous-time signal $\boldsymbol x(\cdot)$,
$\|\boldsymbol x\|_{L_2}:=
(\int_0^\infty\|\boldsymbol x(t)\|^2dt)^{1/2}$, with
$\boldsymbol x\in L_2$ if $\|\boldsymbol x\|_{L_2}<\infty$. 
For a complex matrix $\boldsymbol A$, $\boldsymbol A^*$ denotes its conjugate
transpose, while $|\mathcal S|$ denotes the cardinality of a finite set
$\mathcal S$.
For a stable transfer matrix $\boldsymbol T(s)$, its $\mathcal H_\infty$ norm
is defined as $\|\boldsymbol T\|_\infty
:=
\sup_{\omega\in\mathbb R}
\bar{\sigma}\!\left(\boldsymbol T(j\omega)\right)$,
where $\bar{\sigma}(\cdot)$ denotes the maximum singular value. Equivalently,
for the zero-initial-state input--output relation
$\boldsymbol Z(s)=\boldsymbol T(s)\boldsymbol U(s)$,
$\|\boldsymbol T\|_\infty
=
\sup_{\substack{\boldsymbol u\in {L_2}\\\boldsymbol u\neq\boldsymbol 0}}
\frac{\|\boldsymbol z\|_{L_2}}
{\|\boldsymbol u\|_{L_2}}$.

\begin{table}[htbp]
\centering
\vspace{-0.45cm}
\caption{Selected notation used throughout the paper.}
\label{tab:notation}
\footnotesize
\setlength{\tabcolsep}{3pt}
\renewcommand{\arraystretch}{0.95}
\begin{tabular}{lll}
\hline
Symbol & Meaning & Introduced in \\
\hline
$\boldsymbol T$ & DHS temperature & Eq.~\eqref{dyma1} \\
$\boldsymbol x$ & augmented DHS state & Eq.~\eqref{augdhs_nominal} \\
$\boldsymbol \xi$ & integral state & Eq.~\eqref{augdhs_nominal} \\
$s$ & Laplace variable & Sec.~\ref{subsec:multichannel_response} \\
$\boldsymbol w^h$ & heat-side disturbance & Eq.~\eqref{origindy} \\
$\boldsymbol w_q$ & combined disturbance & Sec.~\ref{subsec:multichannel_response} \\
$\boldsymbol r_i,\boldsymbol\nu_i,\boldsymbol z_i$ & channel state/input/output & Eq.~\eqref{channelpre} \\
$\boldsymbol G_f,\boldsymbol G_d,\boldsymbol G_h$ & grid-facing channels & Eq.~\eqref{threeportsfdh} \\
$\boldsymbol G_G,\boldsymbol G_T$ & DHS-regulation channels & Eq.~\eqref{twoportsfdh} \\
\hline
\end{tabular}
\end{table}
\vspace{-0.5cm}

\subsection{Heat-pump interface}

Let $\mathcal E_{HP}$ and $\mathcal H_{HP}$ denote HP-connected EPS
buses and DHS edges, respectively, with HP $i$ connected to
$k_i\in\mathcal E_{HP}$ and $j_i\in\mathcal H_{HP}$. The gains
$\gamma_i^E>0,\gamma_i^H>0$ and coefficient of performance $CoP_i$ are
constant over the frequency-support time scale. The HP electrical demand and
heat injection are
\begin{subequations}\label{firstorder}
\begin{align}
p_{k_i}^H
&=
\underbrace{\gamma_i^E\omega_{k_i}}_{\text{direct frequency response}}
+
\underbrace{\gamma_i^H S_i^H}_{\text{DHS-state-dependent contribution}},
\label{heatpumpcopcal}\\
q_{j_i}^P
&=
CoP_i\,p_{k_i}^H,
\label{heatpumpcop}
\end{align}
\end{subequations}
where 
$S_i^H$ is generated from the DHS thermal states and calculated as~\eqref{eq:coord_signal_nominal}.
The two terms in~\eqref{heatpumpcopcal} represent the frequency-responsive component
and DHS-state-dependent contribution, respectively. The frequency-responsive term defines the desired grid-facing
response, while the DHS-state-dependent contribution exposes slower thermal
dynamics at the HP electrical interface. Dynamic-port shaping targets
the latter while preserving the former during secondary frequency control.

\subsection{Electric power system}

\subsubsection{Frequency dynamics of electric buses}

Buses $i\in\mathcal E_{DG}$ host controllable generators, while
$i\in\mathcal E_{HP}$ additionally host HPs, with
$\mathcal E=\mathcal E_{DG}\cup\mathcal E_{HP}$. At bus $i$, $P_i$ denotes
net outgoing power, while $\omega_i,D_i,P_i^G,u_i^G,P_i^L,M_i$ denote
frequency deviation, damping, controllable generation, generation-control
input, incremental load, and inertia, respectively.
To capture generation-response dynamics, the controllable generators at each bus $i$ are represented by an aggregate first-order model,
\begin{equation}\label{gen_dyn}
T_{g,i}\dot P_i^G=-P_i^G+u_i^G,\qquad i\in\mathcal E,
\end{equation}
where $T_{g,i}>0$ is the effective turbine--governor time constant.

With $p_i^H$ given by \eqref{heatpumpcopcal}, the bus frequency dynamics are
\begin{subequations}\label{busfre1}
\begin{align}
M_i\dot\omega_i
&=-D_i\omega_i-P_i+P_i^G,
&& i\in\mathcal E_{DG}, \label{busfrea}\\
M_i\dot\omega_i
&=-D_i\omega_i-P_i+P_i^G-p_i^H,
&& i\in\mathcal E_{HP}. \label{hpdhs}
\end{align}
\end{subequations}
The network power-flow and
angle dynamics are
\begin{subequations}\label{busfre2}
\begin{align}
P_i
&=\sum_j |B_{ij}|V_iV_j\sin(\theta_i-\theta_j)+P_i^L,
\qquad i\in\mathcal E, \label{busfre2a}\\
\dot{\boldsymbol x}_\theta
&=\boldsymbol R_I\boldsymbol\omega, \label{busfre2b}
\end{align}
\end{subequations}
The network is connected, with $M_i>0$ and $D_i\geq0$. Here $n_{\mathcal E}:=|\mathcal E|$, $\boldsymbol x_\theta$ is the phase-angle deviation vector with the
reference-bus angle removed, $\boldsymbol\omega$ is the vector of bus-frequency
deviations, and
$\boldsymbol R_I=
[\boldsymbol I_{n_{\mathcal E}-1},-\boldsymbol 1_{n_{\mathcal E}-1}]$. 

\subsubsection{Primary and secondary frequency control}

The controllable generation input consists of primary frequency control and secondary frequency control through automatic generation control (AGC) with the primary-control droop and AGC
integral gains $K_i^P>0$ and $K^I>0$:
\begin{subequations}\label{eq:primary_secondary_control}
\begin{align}
u_i^G
&=-K_i^P\omega_i+\beta_iK^Ig,
\qquad i\in\mathcal E, \label{eq:generation_control}\\
\dot g
&=-\boldsymbol\alpha^\top\boldsymbol\omega,
\label{eq:agc_state}
\end{align}
\end{subequations}
where the nonnegative vectors $\boldsymbol\alpha$ and
$\boldsymbol\beta$ satisfy
$\boldsymbol 1^\top\boldsymbol\alpha
=
\boldsymbol 1^\top\boldsymbol\beta
=1$.
Droop provides primary frequency response and AGC restores nominal frequency on a
slower time scale. The HP responds to local frequency deviation throughout
restoration; AGC acts through controllable generation.

\begin{assumption}[EPS equilibrium]\label{ass:eps_equilibrium} For any $\boldsymbol{\bar{P}^L}$ and  $\boldsymbol{\bar{p}^H}$, the EPS~(\ref{gen_dyn}-\ref{busfre2}) with controller~\eqref{eq:primary_secondary_control} admits an equilibrium $(\boldsymbol{x_\theta^\star}, \boldsymbol{\omega^\star}, \boldsymbol{P^{G\star}}, \boldsymbol{P^\star},\boldsymbol{u^{G\star}},g^\star)$,  
where $\boldsymbol{\omega^\star} = \omega^{com}\boldsymbol{1}_{n_{\mathcal{E}}}$ and $\omega^{com}$ is the synchronized EPS frequency. Moreover, the equilibrium angle differences satisfy $|\theta_i^\star-\theta_j^\star|<\pi/2$ for all connected buses $i$ and $j$. \end{assumption}

\begin{lemma}\label{lem:eps_frequency_restoration} Under Assumption~\ref{ass:eps_equilibrium}, for any constant disturbances 
$\boldsymbol{\bar{P}^{L}}$ and $\boldsymbol{\bar{p}^{H}}$, every equilibrium of the closed-loop EPS \eqref{gen_dyn}--\eqref{busfre2} with controller \eqref{eq:primary_secondary_control} satisfies $\boldsymbol{\omega^{\star}=0}$. \end{lemma}

\begin{assumption}[Input-strict shifted passivity of the generation-side dynamics]
\label{ass:net_supply_passivity}
For each equilibrium of the closed-loop EPS, define
$\tilde{\boldsymbol u}_g:=-\boldsymbol\omega,
\quad
\tilde{\boldsymbol y}_g
:=
\boldsymbol P^G-\boldsymbol P^{G\star}$.
The generation--governor--AGC dynamics~\eqref{gen_dyn},\eqref{eq:primary_secondary_control} are input-strictly shifted passive~\cite{kasis2016primary} from
$\tilde{\boldsymbol u}_g$ to $\tilde{\boldsymbol y}_g$. Specifically, there
exist a nonnegative equilibrium-referenced storage function $V_g$ and a
constant $\rho_g>0$ such that $\dot V_g
\le
\tilde{\boldsymbol u}_g^\top\tilde{\boldsymbol y}_g
-\rho_g\|\tilde{\boldsymbol u}_g\|^2$.
Higher-order generation and control models satisfying the condition can be included without changing the
subsequent port-based interconnection analysis.
\end{assumption}

\subsubsection{EPS port and passivity}

From the EPS perspective, the HP electrical demand $\boldsymbol p^H$ is an
input-port signal generated by the HP-connected DHS.

\begin{definition}[EPS port variables]\label{def:eps_port}
Define $n_{HP}:=|\mathcal H_{HP}|$. Let $\boldsymbol S_{HP}\in\mathbb R^{n_{\mathcal E}\times n_{HP}}$
denote the HP-bus selection matrix, whose $i$th column is the standard basis
vector associated with bus $k_i$. The vector of frequency deviations at the
HP buses is $\boldsymbol\omega^{HP}
=
\boldsymbol S_{HP}^{\top}\boldsymbol\omega$.
The EPS port variables are defined as
\begin{equation}\notag
\boldsymbol u_1:=-\boldsymbol p^H,
\qquad
\boldsymbol y_1:=\boldsymbol\omega^{HP}
=\boldsymbol S_{HP}^{\top}\boldsymbol\omega .
\end{equation}
\end{definition}

\begin{lemma}[Output-strict passivity of the EPS port]\label{lem:eps_passive}
Under Assumptions~\ref{ass:eps_equilibrium} and \ref{ass:net_supply_passivity}, consider the EPS model
\eqref{gen_dyn}--\eqref{busfre2} with controller
\eqref{eq:primary_secondary_control}. With the equilibrium-referenced deviation $\tilde{\boldsymbol u}_1 := -(\boldsymbol p^H-\boldsymbol p^{H\star})$ of the port input $\boldsymbol u_1$ in Definition~\ref{def:eps_port}
and the output $\boldsymbol{y_1 := \omega^{HP}}$, the EPS is locally strictly output passive from $\tilde{\boldsymbol u}_1$ to
$\boldsymbol{y_1}$: there exist a storage function $V_{\mathrm{e}}$ 
and a constant $\rho_e>0$ such that 
\begin{equation}\notag
\dot V_{\mathrm e}
\le
\boldsymbol y_1^\top\tilde{\boldsymbol u}_1
-\rho_e\|\boldsymbol y_1\|^2 .
\end{equation}
The EPS-port passivity property is independent of the DHS regulator used
to generate $\boldsymbol p^H$.
\end{lemma}

\subsection{District heating system}

\subsubsection{Temperature dynamics}

Following~\cite{qin2025_tpwrs_coupled}, let $\mathcal H^E$ and $\mathcal H^N$
denote the DHS edge and node sets, with
$\mathcal H^E=\mathcal H_G\cup\mathcal H_{HP}\cup\mathcal H_L$ denoting the sets of
conventional-source, HP, and load edges. For edge $j$, $s(j)$ denotes its
upstream node, and $\mathcal T_k$ the set of edges entering node $k$. Their
temperature dynamics are

{\small
\begin{subequations}\label{temdy}
\begin{align}
\rho C_p V_j^E\dot T_j^E
&=\rho C_p\phi_j^E(T_{s(j)}^N-T_j^E)+q_j^G+q_j^P-q_j^L,
j\in\mathcal H^E, \label{eq:edge_temp_dyn}\\
\rho C_p V_k^N\dot T_k^N
&=\sum_{j\in\mathcal T_k}\rho C_p\phi_j^E(T_j^E-T_k^N),
k\in\mathcal H^N, \label{eq:node_temp_dyn}
\end{align}
\end{subequations}
}

\noindent Here $T^E,T^N$ are outlet temperatures, $V^E,V^N$ are volumes, $\phi^E$ is
volumetric flow, and $\rho,C_p$ are the density and specific heat of water.
$q^G,q^P,q^L$ denote conventional-source injection, HP injection from
\eqref{heatpumpcop}, and load extraction; only the term associated with each
edge type is nonzero.
Stack the temperatures as
$\boldsymbol T=[(\boldsymbol T^G)^\top,(\boldsymbol T^{HP})^\top,(\boldsymbol T^L)^\top,(\boldsymbol T^N)^\top]^\top\in\mathbb R^{n_T}$,
where $n_T:=|\mathcal H^E|+|\mathcal H^N|$, and define
$\boldsymbol V:=\mathrm{diag}(\{V_j^E\}_{j\in\mathcal H^E},\{V_k^N\}_{k\in\mathcal H^N})$.
Using the Kirchhoff matrix $\boldsymbol A_h$ in~\cite{qin2025_tpwrs_coupled},
\eqref{temdy} can be written as
\begin{equation}\label{dyma1}
\rho C_p\boldsymbol V\dot{\boldsymbol T}
=-\rho C_p \boldsymbol A_h\boldsymbol T
+\boldsymbol E_G\boldsymbol q^G
+\boldsymbol E_{HP}\boldsymbol q^P
-\boldsymbol E_L\boldsymbol q^L,
\end{equation}
where
$\boldsymbol E_G=[\boldsymbol I,\boldsymbol 0,\boldsymbol 0,\boldsymbol 0]^\top$,
$\boldsymbol E_{HP}=[\boldsymbol 0,\boldsymbol I,\boldsymbol 0,\boldsymbol 0]^\top$,
and $\boldsymbol E_L=[\boldsymbol 0,\boldsymbol 0,\boldsymbol I,\boldsymbol 0]^\top$.
Define
{\small
\begin{equation}\notag
\begin{aligned}
\boldsymbol A&:=\boldsymbol V^{-1}\boldsymbol A_h,\qquad
\boldsymbol B_G:=(\rho C_p)^{-1}\boldsymbol V^{-1}\boldsymbol E_G,\\
\boldsymbol B_P&:=(\rho C_p)^{-1}\boldsymbol V^{-1}\boldsymbol E_{HP},
\qquad
\boldsymbol w^h:=-(\rho C_p)^{-1}\boldsymbol V^{-1}\boldsymbol E_L\boldsymbol q^L.
\end{aligned}
\end{equation}
}
Then \eqref{dyma1} becomes
\begin{equation}\label{origindy}
\dot{\boldsymbol T}
=-\boldsymbol A\boldsymbol T+\boldsymbol B_G\boldsymbol q^G
+\boldsymbol B_P\boldsymbol q^P+\boldsymbol w^h .
\end{equation}

\subsubsection{Energy-sharing steady-state objectives}

The steady-state economic dispatch for the DHS under given
$\bar{\boldsymbol q}^P$ and $\bar{\boldsymbol w}^h$ is modeled as two
optimization problems:
\begin{subequations}\label{opt1_nominal_all}
\begin{align}
\textbf{OPT-1:}
\min_{\boldsymbol q^G\in\mathbb R^{n_G},\,\boldsymbol T\in\mathbb R^{n_T}}
J_1
&=\frac12\boldsymbol q^{G\top}\boldsymbol W^G\boldsymbol q^G,
\label{opt1_nominal_obj}\\
\text{s.t.}\quad
\boldsymbol A\boldsymbol T
&=\boldsymbol B_G\boldsymbol q^G+\boldsymbol B_P\bar{\boldsymbol q}^{P}
+\bar{\boldsymbol w}^h,
\label{opt1_nominal_con}
\end{align}
\end{subequations}
where $n_G:=|\mathcal H_G|$, $J_1$ denotes the conventional-source heat
generation cost, and $\boldsymbol W^G=\mathrm{diag}(W_i^G)\succ0$. The
temperature deviation cost is minimized in
\begin{subequations}\label{opt2_nominal_all}
\begin{align}
\textbf{OPT-2:}
\min_{\boldsymbol T\in\mathbb R^{n_T},\,\alpha\in\mathbb R}
J_2&=\frac12\boldsymbol T^\top\boldsymbol W^T\boldsymbol T,
\label{opt2_nominal_obj}\\
\text{s.t.}\quad
\boldsymbol T
&=\boldsymbol A^\dagger(\boldsymbol B_G\boldsymbol q^{G\star}
+\boldsymbol B_P\bar{\boldsymbol q}^{P}+\bar{\boldsymbol w}^h)
+\alpha\boldsymbol 1,
\label{opt2_nominal_con}
\end{align}
\end{subequations}
where $\boldsymbol A^\dagger$ is the Moore--Penrose pseudoinverse of
$\boldsymbol A$, and $\boldsymbol W^T=\mathrm{diag}(W_i^T)\succ0$ is the
temperature-deviation penalty matrix.

\begin{definition}[Conventional-source marginal-cost matrix]\label{def:WGm_nominal}
Let $\boldsymbol R_G:=[\boldsymbol I_{n_G-1},-\boldsymbol 1_{n_G-1}]$. Define
$\boldsymbol W_G^M:=\boldsymbol R_G\boldsymbol W^G$.
\end{definition}

\begin{theorem}\label{thm:dhs_opt_nominal}
For a connected, flow-conserving DHS, if (\ref{origindy}) achieves equilibrium at $\boldsymbol{{T^*}}$ and $\boldsymbol{{q^G}^*}$, and satisfies
\begin{equation}\label{eq:nominal_stationarity_conditions}
\boldsymbol W_G^M\boldsymbol q^{G\star}=\boldsymbol 0,
\qquad
\boldsymbol 1_{n_T}^\top\boldsymbol W^T\boldsymbol T^\star=0,
\end{equation}
then $\boldsymbol q^{G\star}$ is the unique optimizer of \textbf{OPT-1} in
\eqref{opt1_nominal_all}, and $\boldsymbol T^\star$ is the unique optimizer of
\textbf{OPT-2} in \eqref{opt2_nominal_all}.
\end{theorem}

Based on the temperature-selection condition in
\eqref{eq:nominal_stationarity_conditions}, the DHS regulation signal is chosen
without requiring heat-disturbance prediction as
\begin{equation}\label{eq:coord_signal_nominal}
S_i^H=\boldsymbol 1_{n_T}^\top \boldsymbol W^T\boldsymbol T,\quad i=1,\ldots,n_{HP}.
\end{equation}

\subsubsection{Augmented DHS dynamics}

Substituting \eqref{eq:coord_signal_nominal} and \eqref{firstorder} into
\eqref{origindy} gives
$\dot{\boldsymbol T}
=\boldsymbol A_c\boldsymbol T+\boldsymbol B_G\boldsymbol q^G
+\boldsymbol B_\omega\boldsymbol\omega^{HP}+\boldsymbol w^h$,
where $\boldsymbol A_c=-\boldsymbol A+
\boldsymbol B_P\boldsymbol{CoP}\boldsymbol\gamma^H
\boldsymbol 1_{n_{HP}}\boldsymbol 1_{n_T}^\top\boldsymbol W^T$,
$\boldsymbol B_\omega=\boldsymbol B_P\boldsymbol{CoP}\boldsymbol\gamma^E$,
$\boldsymbol{CoP}:=\mathrm{diag}(CoP_i)$, and
$\boldsymbol\gamma^E$, $\boldsymbol\gamma^H$ are diagonal matrices containing
the HP coefficients in \eqref{firstorder}.

To embed the optimality conditions in \eqref{eq:nominal_stationarity_conditions},
define
\begin{equation}\label{eq:error_nominal}
\boldsymbol e=\boldsymbol C_e\boldsymbol T+\boldsymbol D_e\boldsymbol q^G=\begin{bmatrix}
    \boldsymbol{e}_G\\e_T
\end{bmatrix},
\end{equation}
where
$\boldsymbol C_e=\begin{bmatrix}
\boldsymbol 0_{(n_G-1)\times n_T}\\
\boldsymbol 1_{n_T}^\top\boldsymbol W^T
\end{bmatrix}$,
$\boldsymbol D_e=\begin{bmatrix}
\boldsymbol W_G^M\\
\boldsymbol 0_{1\times n_G}\end{bmatrix}$.

Introducing the integral state $\dot{\boldsymbol\xi}=\boldsymbol e$ and letting
$\boldsymbol x=[\boldsymbol T^\top,\boldsymbol\xi^\top]^\top$, the augmented
DHS model is
\begin{equation}\label{augdhs_nominal}
\dot{\boldsymbol x}
=\boldsymbol{\hat A}\boldsymbol x+
\boldsymbol{\hat B}\boldsymbol q^G+
\boldsymbol{\hat B}_\omega\boldsymbol\omega^{HP}+
\boldsymbol{\hat B}_h\boldsymbol w^h,
\end{equation}
with
$\boldsymbol{\hat A}=\begin{bmatrix}\boldsymbol A_c&\boldsymbol 0\\
\boldsymbol C_e&\boldsymbol 0\end{bmatrix}$,
$\boldsymbol{\hat B}=\begin{bmatrix}\boldsymbol B_G\\\boldsymbol D_e\end{bmatrix}$,
$\boldsymbol{\hat B}_\omega=\begin{bmatrix}\boldsymbol B_\omega\\\boldsymbol 0\end{bmatrix}$,
$\boldsymbol{\hat B}_h=\begin{bmatrix}\boldsymbol I\\\boldsymbol 0\end{bmatrix}$.

\subsubsection{DHS port representation under augmented-state feedback}

The augmented DHS is closed with the feedback regulator
$\boldsymbol q^G=-\boldsymbol K\boldsymbol x$.
The DHS-side port input is the HP-bus frequency deviation supplied by the EPS,
and the port output is the HP electrical-power consumption returned to the EPS.

\begin{definition}[Augmented DHS port variables]\label{def:dhs_port}
The DHS port variables are defined as
$\boldsymbol u_2:=\boldsymbol\omega^{HP}$, $\boldsymbol y_2:=\boldsymbol p^H$.
\end{definition}

The DHS port output is
\begin{equation}\label{eq:y2_nominal}
\boldsymbol y_2=\boldsymbol p^H
=\boldsymbol\gamma^E\boldsymbol\omega^{HP}+\boldsymbol C_0\boldsymbol x,
\end{equation}
where $\boldsymbol C_0=\boldsymbol\gamma^H
\begin{bmatrix}
\boldsymbol 1_{n_{HP}}\boldsymbol 1_{n_T}^\top\boldsymbol W^T &
\boldsymbol 0
\end{bmatrix}$.
With
\begin{equation}\label{eq:Acl_nominal_def}
\boldsymbol{\hat A}_{cl}(\boldsymbol K):=
\boldsymbol{\hat A}-\boldsymbol{\hat B}\boldsymbol K,
\end{equation}
the closed-loop augmented DHS is
\begin{equation}\label{closeloopdhs_nominal}
\dot{\boldsymbol x}
=\boldsymbol{\hat A}_{cl}(\boldsymbol K)\boldsymbol x
+\boldsymbol{\hat B}_\omega\boldsymbol\omega^{HP}
+\boldsymbol{\hat B}_h\boldsymbol w^h .
\end{equation}

\begin{lemma}\label{lem:dhs_cl_properties}
Suppose that $\boldsymbol{\hat A}_{cl}(\boldsymbol K)$ is exponentially stable.
Then the following properties hold.

\noindent\textit{(i)} For any constant heat disturbance $\bar{\boldsymbol w}^h$
and any constant HP-bus frequency input $\bar{\boldsymbol\omega}^{HP}$, the
feedback DHS \eqref{closeloopdhs_nominal} admits a unique equilibrium
\begin{equation}\label{eq:dhs_equilibrium_xstar}
\boldsymbol x^\star
=-\boldsymbol{\hat A}_{cl}(\boldsymbol K)^{-1}
\left(\boldsymbol{\hat B}_\omega\bar{\boldsymbol\omega}^{HP}
+\boldsymbol{\hat B}_h\bar{\boldsymbol w}^h\right).
\end{equation}
Moreover, every trajectory satisfies $\boldsymbol x(t)\to\boldsymbol x^\star$
and $\boldsymbol e(t)\to\boldsymbol 0$.

\noindent\textit{(ii)} At the equilibrium in \eqref{eq:dhs_equilibrium_xstar},
the DHS-state-dependent contribution of the HP electrical-power output \eqref{eq:y2_nominal} has zero
steady-state value:
\begin{equation}\label{eq:C0_xstar_zero}
\boldsymbol C_0\boldsymbol x^\star=\boldsymbol 0.
\end{equation}
Consequently,
\begin{equation}\label{eq:C0_Acl_inverse_zero}
\boldsymbol C_0\boldsymbol{\hat A}_{cl}(\boldsymbol K)^{-1}\boldsymbol{\hat B}_\omega
=\boldsymbol 0,
\qquad
\boldsymbol C_0\boldsymbol{\hat A}_{cl}(\boldsymbol K)^{-1}\boldsymbol{\hat B}_h
=\boldsymbol 0 .
\end{equation}

\noindent\textit{(iii)} With respect to the equilibrium
$(\boldsymbol x^\star,\boldsymbol u_2^\star,\boldsymbol y_2^\star)$, define
\[
\tilde{\boldsymbol x}:=\boldsymbol x-\boldsymbol x^\star,\qquad
\tilde{\boldsymbol u}_2:=\boldsymbol u_2-\boldsymbol u_2^\star,\qquad
\tilde{\boldsymbol y}_2:=\boldsymbol y_2-\boldsymbol y_2^\star.
\]
The corresponding deviation dynamics of
\eqref{eq:y2_nominal} and \eqref{closeloopdhs_nominal} are
\begin{equation}\label{eq:dev_dyn_shifted_A}
\begin{aligned}
\dot{\tilde{\boldsymbol x}}
&=
\boldsymbol{\hat A}_{cl}(\boldsymbol K)\tilde{\boldsymbol x}
+\boldsymbol{\hat B}_\omega\tilde{\boldsymbol u}_2,\\
\tilde{\boldsymbol y}_2
&=
\boldsymbol C_0\tilde{\boldsymbol x}
+\boldsymbol\gamma^E\tilde{\boldsymbol u}_2.
\end{aligned}
\end{equation}
Moreover, if $\tilde{\boldsymbol u}_2(t)\equiv\boldsymbol 0$, the only
zero-input equilibrium is
$\tilde{\boldsymbol x}=\boldsymbol 0$.
\end{lemma}

\section{Port-Interconnection Analysis}
\label{sec:nominal_synthesis}

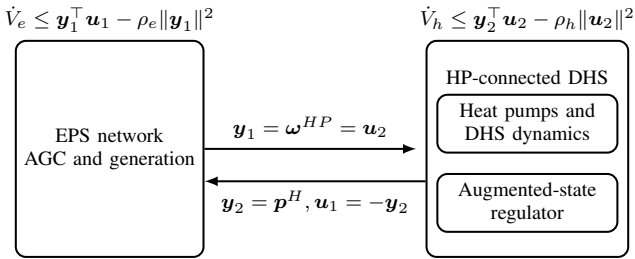
\begin{figure}[htbp]
\centering
\resizebox{0.98\columnwidth}{!}{%
\begin{tikzpicture}[
    outerbox/.style={draw,rounded corners,minimum width=2.8cm,minimum height=3.0cm,align=center,font=\footnotesize},
    innerbox/.style={draw,rounded corners,minimum width=2.5cm,minimum height=0.8cm,align=center,font=\footnotesize},
    leftbox/.style={draw,rounded corners,minimum width=2.3cm,minimum height=3.0cm,align=center,font=\footnotesize},
    >=latex, thick]
\node[leftbox] (eps) at (0,0) {EPS network\\AGC and generation};
\node[outerbox] (rhs) at (5.8,0) {};
\node[font=\footnotesize] at (5.8,1.0) {HP-connected DHS};
\node[innerbox] (hp) at (5.8,0.35) {Heat pumps and\\DHS dynamics};
\node[innerbox] (dhs) at (5.8,-0.75) {Augmented-state\\regulator};
\draw[->] (eps.east) -- node[above,font=\footnotesize]
{$\boldsymbol y_1={\boldsymbol\omega}^{HP}={\boldsymbol u}_2$}
([xshift=-0.13cm]rhs.west);
\draw[->] ([yshift=-0.45cm]rhs.west) -- node[below,font=\footnotesize]
{${\boldsymbol y}_2={\boldsymbol p}^{H},{\boldsymbol u}_1=-{\boldsymbol y}_2$}
([yshift=-0.45cm]eps.east);
\node[font=\footnotesize, align=center] at (0,1.8)
{$\dot V_e \le {\boldsymbol y}_1^\top {\boldsymbol u}_1
-\rho_e\|{\boldsymbol y}_1\|^2$};
\node[font=\footnotesize, align=center] at (5.8,1.8)
{$\dot V_h \le {\boldsymbol y}_2^\top {\boldsymbol u}_2
-\rho_h\|{\boldsymbol u}_2\|^2$};
\end{tikzpicture}
}
\caption{EPS--HP--DHS port interconnection used in the
passivity proof.}
\label{fig:eps_dhs_port_interconnection}
\end{figure}

This section establishes the equilibrium and stability properties of the
EPS--HP--DHS interconnection and derives tractable matrix-inequality
conditions ensuring passivity of the grid-facing DHS port and exponential
stability of the augmented DHS dynamics. These conditions serve as common
feasibility constraints for the dynamic-port shaping designs in
Section~\ref{sec:loop_shaping}. The interconnection is shown in
Fig.~\ref{fig:eps_dhs_port_interconnection}.

\subsection{Equilibrium and port-interconnection stability}

\begin{lemma}[Target equilibrium of the closed-loop EPS--HP--DHS  system]\label{lem:nominal_coupled_equilibrium}
Suppose Assumptions~\ref{ass:eps_equilibrium} and~\ref{ass:net_supply_passivity}
hold and $\boldsymbol{\hat A}_{cl}(\boldsymbol K)$ in \eqref{eq:Acl_nominal_def}
is exponentially stable. Then, for any constant electrical disturbance
$\bar{\boldsymbol P}^L$ and heat disturbance $\bar{\boldsymbol w}^h$, the
closed-loop EPS--HP--DHS system \eqref{gen_dyn}--\eqref{eq:primary_secondary_control},
\eqref{closeloopdhs_nominal} admits an equilibrium satisfying
$\boldsymbol\omega^\star=\boldsymbol 0$,
$\boldsymbol e^\star=\boldsymbol 0$.
At this equilibrium, the EPS frequency-restoration objective in
Lemma~\ref{lem:eps_frequency_restoration} and the DHS steady-state optimality
conditions in Theorem~\ref{thm:dhs_opt_nominal} are satisfied.
\end{lemma}

\begin{theorem}[Local asymptotic stability of the interconnected equilibrium]\label{thm:nominal_interconnection_stability}
Under Assumptions~\ref{ass:eps_equilibrium} and~\ref{ass:net_supply_passivity},
if the augmented DHS in~\eqref{closeloopdhs_nominal},\eqref{eq:y2_nominal} is input-strict passive with respect to the input--output pair $(\boldsymbol{u_2 = \omega^{HP}},\, \boldsymbol{y_2 = p^H})$, and if $\boldsymbol{\hat A}_{cl}$ is Hurwitz, with locally positive-definite storage, then the equilibrium defined in Lemma~\ref{lem:nominal_coupled_equilibrium} is locally asymptotically stable.
\end{theorem}

\subsection{Passivity and stability synthesis}

Theorem~\ref{thm:nominal_interconnection_stability} requires the augmented DHS
\eqref{closeloopdhs_nominal}, \eqref{eq:y2_nominal} to be input-strictly
passive and $\boldsymbol{\hat A}_{cl}(\boldsymbol K)$ to be Hurwitz. 
For the DHS port in~\eqref{closeloopdhs_nominal} and \eqref{eq:y2_nominal},
use equilibrium-referenced deviations for constant heat disturbances and consider the storage function $V_h(\boldsymbol x)
=
\frac{1}{2}\boldsymbol x^\top\boldsymbol H\boldsymbol x,
\boldsymbol H\succ\boldsymbol 0$.
Input-strict passivity with $\rho_h>0$ requires
$\dot V_h
\leq
\boldsymbol y_2^\top\boldsymbol u_2
-
\rho_h\|\boldsymbol u_2\|^2$, which gives
{\scriptsize \begin{equation}\notag
\begin{bmatrix}
\boldsymbol x\\
\boldsymbol u_2
\end{bmatrix}^{\!\top}
\begin{bmatrix}
\boldsymbol{\hat A}_{cl}(\boldsymbol K)^\top\boldsymbol H
+
\boldsymbol H\boldsymbol{\hat A}_{cl}(\boldsymbol K)
&
\boldsymbol H\boldsymbol{\hat B}_{\omega}
-
\boldsymbol C_0^\top
\\
*
&
-\bigl(
\boldsymbol\gamma^E
+
\boldsymbol\gamma^{E\top}
\bigr)
+
2\rho_h\boldsymbol I
\end{bmatrix}
\begin{bmatrix}
\boldsymbol x\\
\boldsymbol u_2
\end{bmatrix}
\leq 0.
\end{equation}}
Hence, a sufficient KYP condition for input-strict passivity is
\begin{equation}\label{eq:KYP_BMI_shifted_A}
\begin{bmatrix}
\boldsymbol{\hat A}_{cl}(\boldsymbol K)^\top\boldsymbol H
+
\boldsymbol H\boldsymbol{\hat A}_{cl}(\boldsymbol K)
&
\boldsymbol H\boldsymbol{\hat B}_{\omega}
-
\boldsymbol C_0^\top
\\
*
&
-\bigl(
\boldsymbol\gamma^E
+
\boldsymbol\gamma^{E\top}
\bigr)
+
2\rho_h\boldsymbol I
\end{bmatrix}
\preceq
\boldsymbol 0.
\end{equation}
Exponential stability with a prescribed decay rate $\beta>0$ is imposed
through
\begin{equation}\label{eq:decay_rate_BMI}
\boldsymbol{\hat A}_{cl}(\boldsymbol K)^\top\boldsymbol H
+\boldsymbol H\boldsymbol{\hat A}_{cl}(\boldsymbol K)
+2\beta\boldsymbol H
\preceq\boldsymbol 0,
\end{equation}
which guarantees
$\operatorname{Re}(\lambda(\boldsymbol{\hat A}_{cl}(\boldsymbol K)))
\leq-\beta$. Conditions \eqref{eq:KYP_BMI_shifted_A} and
\eqref{eq:decay_rate_BMI} are bilinear in $\boldsymbol H$ and
$\boldsymbol K$.

Let $\boldsymbol X:=\boldsymbol H^{-1}\succ\boldsymbol 0,\quad
\boldsymbol Y:=\boldsymbol K\boldsymbol X$.
Applying a congruence transformation $\operatorname{diag} \bigl( \boldsymbol X, \boldsymbol I_{n_{HP}} \bigr)$\footnote{The detailed transformation is
given in the Appendix.} converts \eqref{eq:KYP_BMI_shifted_A} into
\begin{equation}\label{eq:LMI_shifted}
\boldsymbol\Psi^{\rm p}:=
\begin{bmatrix}
\boldsymbol M_{11} & \boldsymbol M_{12}\\
\boldsymbol M_{12}^\top &
-\bigl(\boldsymbol\gamma^E+\boldsymbol\gamma^{E\top}\bigr)
+2\rho_h\boldsymbol I
\end{bmatrix}
\preceq\boldsymbol 0,
\end{equation}
where
$\boldsymbol M_{11}:=
\operatorname{sym}\!\left(
\boldsymbol{\hat A}\boldsymbol X
-\boldsymbol{\hat B}\boldsymbol Y
\right),\quad
\boldsymbol M_{12}
:=
\boldsymbol{\hat B}_\omega
-\boldsymbol X\boldsymbol C_0^\top$.
Similarly, \eqref{eq:decay_rate_BMI} becomes
\begin{equation}\label{eq:decay_rate_LMI_short}
\boldsymbol M_{11}
\preceq-2\beta\boldsymbol X,
\qquad \beta>0.
\end{equation}
For prescribed $\beta$ and $\rho_h$, \eqref{eq:LMI_shifted} and
\eqref{eq:decay_rate_LMI_short} are LMIs in
$(\boldsymbol X,\boldsymbol Y)$, and the feedback gain is recovered as
$\boldsymbol K=\boldsymbol Y\boldsymbol X^{-1}$.

\subsection{Passivity-Feasible Baseline}

For prescribed passivity margin $\rho_h>0$ and decay rate $\beta>0$, a
passivity-feasible baseline is obtained from the feasibility problem
\begin{equation}\label{eq:baseline_passivity_synthesis}
\begin{aligned}
\mathrm{find}\quad
&\boldsymbol X,\boldsymbol Y\\
\mathrm{s.t.}\quad
&\boldsymbol X\succ\boldsymbol 0,\quad
\boldsymbol\Psi^{\rm p}\preceq\boldsymbol 0,\quad
\boldsymbol M_{11}\preceq-2\beta\boldsymbol X.
\end{aligned}
\end{equation}
The controller is then recovered as
$\boldsymbol K_{\rm base}
=
\boldsymbol Y\boldsymbol X^{-1}$.
This design provides the minimum requirements for the
EPS--HP--DHS interconnection: input-strict passivity of the DHS port and
exponential stability of the augmented DHS dynamics.

\section{$\mathcal H_\infty$ Loop-Shaping Synthesis}
\label{sec:loop_shaping}

The passivity and stability conditions in Section~\ref{sec:nominal_synthesis}
guarantee EPS--HP--DHS stability but do not constrain how
DHS-state-dependent dynamics and heat-side disturbances appear at the
electrical interface, nor the associated DHS-regulation dynamics. Multi-channel
$\mathcal H_\infty$ shaping attenuates DHS-state-dependent electrical dynamics,
heat-disturbance propagation, and DHS-regulation errors during secondary
frequency control, while preserving the desired grid-facing response.

\subsection{Multi-channel dynamic response}
\label{subsec:multichannel_response}
\subsubsection{Grid-Facing Channels}
Starting from the pre-disturbance equilibrium, \eqref{closeloopdhs_nominal} and
\eqref{eq:y2_nominal} yield
\begin{equation}\notag
\begin{aligned}
\boldsymbol p^H(s)
=&
\boldsymbol\gamma^E\boldsymbol\omega^{HP}(s)
+
\boldsymbol C_0
\bigl(s\boldsymbol I-\boldsymbol{\hat A}_{cl}(\boldsymbol K)\bigr)^{-1}
\boldsymbol{\hat B}_\omega
\boldsymbol\omega^{HP}(s)
\\
&+
\boldsymbol C_0
\bigl(s\boldsymbol I-\boldsymbol{\hat A}_{cl}(\boldsymbol K)\bigr)^{-1}
\boldsymbol{\hat B}_h
\boldsymbol w^h(s).
\end{aligned}
\end{equation}
Here $s$ denotes the Laplace variable. Define the full frequency-response, frequency-induced dynamic, and heat-disturbance
channels as
\begin{subequations}\label{threeportsfdh}
\begin{align}
\boldsymbol G_f(s;\boldsymbol K)
&:=
\boldsymbol\gamma^E+\boldsymbol G_d(s;\boldsymbol K),
\label{portmapfull}\\
\boldsymbol G_d(s;\boldsymbol K)
&:=
\boldsymbol C_0
\bigl(s\boldsymbol I-\boldsymbol{\hat A}_{cl}(\boldsymbol K)\bigr)^{-1}
\boldsymbol{\hat B}_\omega,
\label{portmapdyn}\\
\boldsymbol G_h(s;\boldsymbol K)
&:=
\boldsymbol C_0
\bigl(s\boldsymbol I-\boldsymbol{\hat A}_{cl}(\boldsymbol K)\bigr)^{-1}
\boldsymbol{\hat B}_h .
\label{portmapheat}
\end{align}
\end{subequations}
For a fixed $\boldsymbol K$,
$\boldsymbol p^H(s)=\boldsymbol G_f(s)\boldsymbol\omega^{HP}(s)
+\boldsymbol G_h(s)\boldsymbol w^h(s)$.
Here, $\boldsymbol G_d$ captures the DHS-state-dependent contribution to the
frequency-driven HP response, while $\boldsymbol G_h$ captures heat-disturbance
propagation to the electrical port. Attenuating $\boldsymbol G_d$ limits
DHS-state-dependent electrical dynamics, keeping
$\boldsymbol G_f$ close to the direct frequency-response gain. Attenuating
$\boldsymbol G_h$ limits heat-disturbance propagation to the electrical port.
These are complementary aspects of maintaining a clean grid-facing response
during secondary frequency control. 

By Lemma~\ref{lem:dhs_cl_properties},
$\boldsymbol G_d(0)=\boldsymbol G_h(0)=\boldsymbol 0$ and
$\boldsymbol G_f(0)=\boldsymbol\gamma^E$, so the DHS dynamics shape the
transient HP response without altering the steady-state frequency
gain. In addition, $\boldsymbol G_d$ and $\boldsymbol G_h$ are strictly proper
and therefore vanish at high frequency, yielding
$\boldsymbol G_f(j\omega)\to\boldsymbol\gamma^E$ as
$\omega\to\infty$~\cite{vanantwerp2000tutorial}.

\subsubsection{DHS-Regulation channels}

Starting from the pre-disturbance equilibrium, define the combined disturbance input and its input matrix as $\boldsymbol w_q
:=
\operatorname{col}
\left(
\boldsymbol\omega^{HP},
\boldsymbol w^h
\right)$,
$\boldsymbol B_q
:=
\begin{bmatrix}
\boldsymbol{\hat B}_\omega&
\boldsymbol{\hat B}_h
\end{bmatrix}$.
From \eqref{closeloopdhs_nominal} and \eqref{eq:error_nominal},
\begin{equation}\notag
\begin{aligned}
\boldsymbol e_G(s)
&=
-\boldsymbol W_G^M\boldsymbol K
\bigl(
s\boldsymbol I-\boldsymbol{\hat A}_{cl}(\boldsymbol K)
\bigr)^{-1}
\boldsymbol B_q\boldsymbol w_q(s),\\
e_T(s)
&=
\boldsymbol C_{e_T}
\bigl(
s\boldsymbol I-\boldsymbol{\hat A}_{cl}(\boldsymbol K)
\bigr)^{-1}
\boldsymbol B_q\boldsymbol w_q(s),
\end{aligned}
\end{equation}
where $\boldsymbol C_{e_T}
:=
\begin{bmatrix}
\boldsymbol 1_{n_T}^{\top}\boldsymbol W^T&
\boldsymbol 0_{1\times n_G}
\end{bmatrix}$.
Accordingly, define the generation-sharing and temperature-regulation channels as
\begin{subequations}\label{twoportsfdh}
\begin{align}
\boldsymbol G_G(s;\boldsymbol K)
&:=
-\boldsymbol W_G^M\boldsymbol K
\bigl(
s\boldsymbol I-\boldsymbol{\hat A}_{cl}(\boldsymbol K)
\bigr)^{-1}
\boldsymbol B_q,\label{generationshare}\\
\boldsymbol G_T(s;\boldsymbol K)
&:=
\boldsymbol C_{e_T}
\bigl(
s\boldsymbol I-\boldsymbol{\hat A}_{cl}(\boldsymbol K)
\bigr)^{-1}
\boldsymbol B_q.\label{temperatureregulation}
\end{align}
\end{subequations}
Thus, for a fixed $\boldsymbol K, \boldsymbol e_G(s)=\boldsymbol G_G(s)\boldsymbol w_q(s)$ and
$e_T(s)=\boldsymbol G_T(s)\boldsymbol w_q(s)$.

\subsection{Multi-channel
$\mathcal H_\infty$ Dynamic Port Attenuation}
\label{subsec:dynamic_attenuation}

The multi-channel
$\mathcal H_\infty$ dynamic-attenuation design jointly limits the five channels
$\boldsymbol G_f$, $\boldsymbol G_d$, $\boldsymbol G_h$,
$\boldsymbol G_G$, and $\boldsymbol G_T$. The first three characterize the
grid-facing response to frequency and heat disturbances, while
$\boldsymbol G_G$ and $\boldsymbol G_T$ quantify DHS generation-sharing and
temperature-regulation errors. The regulator $\boldsymbol K$ is synthesized to
reduce their $\mathcal H_\infty$ gains subject to the DHS-port passivity and
decay-rate constraints in Section~\ref{sec:nominal_synthesis}.

All five channels share the closed-loop state matrix
$\boldsymbol{\hat A}_{cl}(\boldsymbol K)$ and differ only in their input and
output matrices. For compactness, channel $i$ is written as
\begin{equation}\label{channelpre}
\begin{aligned}
\dot{\boldsymbol r}_i
&=
\boldsymbol{\hat A}_{cl}(\boldsymbol K)\boldsymbol r_i
+\boldsymbol B_i\boldsymbol\nu_i,\\
\boldsymbol z_i
&=
\left(
\boldsymbol C_i-\boldsymbol D_{u,i}\boldsymbol K
\right)\boldsymbol r_i
+\boldsymbol D_i\boldsymbol\nu_i\\
&=
\boldsymbol C_{K,i}\boldsymbol r_i
+\boldsymbol D_i\boldsymbol\nu_i,
\end{aligned}
\end{equation}
where $\boldsymbol r_i$ denotes the channel-specific realization state, $\boldsymbol\nu_i$ the generic channel input, and $\boldsymbol z_i$ the corresponding channel output, and $\boldsymbol C_{K,i}:= \boldsymbol C_i-\boldsymbol D_{u,i}\boldsymbol K$.
The five channel realizations are summarized as
\begin{equation}\notag
\begin{array}{c|c|c|c|c|c|c}
\hline
i
&\boldsymbol\nu_i
&\boldsymbol G_i
&\boldsymbol B_i
&\boldsymbol C_i
&\boldsymbol D_{u,i}
&\boldsymbol D_i
\\ \hline
f
&\boldsymbol\omega^{HP}
&\boldsymbol G_f~\text{in}~\eqref{portmapfull}
&\boldsymbol{\hat B}_{\omega}
&\boldsymbol C_0
&\boldsymbol 0
&\boldsymbol\gamma^E
\\
d
&\boldsymbol\omega^{HP}
&\boldsymbol G_d~\text{in}~\eqref{portmapdyn}
&\boldsymbol{\hat B}_{\omega}
&\boldsymbol C_0
&\boldsymbol 0
&\boldsymbol 0
\\
h
&\boldsymbol w^h
&\boldsymbol G_h~\text{in}~\eqref{portmapheat}
&\boldsymbol{\hat B}_h
&\boldsymbol C_0
&\boldsymbol 0
&\boldsymbol 0
\\
G
&\boldsymbol w_q
&\boldsymbol G_G~\text{in}~\eqref{generationshare}
&\boldsymbol B_q
&\boldsymbol 0
&\boldsymbol W_G^M
&\boldsymbol 0
\\
T
&\boldsymbol w_q
&\boldsymbol G_T~\text{in}~\eqref{temperatureregulation}
&\boldsymbol B_q
&\boldsymbol C_{e_T}
&\boldsymbol 0
&\boldsymbol 0\\
\hline
\end{array}
\end{equation}

For each channel, the bound
$\|\boldsymbol G_i\|_\infty<\eta_i$, $\eta_i>0$, implies
$\int_0^\infty
\boldsymbol z_i^\top\boldsymbol z_i\,\mathrm{d}t
<
\eta_i^2
\int_0^\infty
\boldsymbol\nu_i^\top\boldsymbol\nu_i\,\mathrm{d}t$
for every nonzero $\boldsymbol\nu_i\in L_2$ under zero initial state.
Using the quadratic storage function
$V_i(\boldsymbol r_i)=\boldsymbol r_i^\top\boldsymbol H\boldsymbol r_i$,
$\boldsymbol H\succ\boldsymbol 0$,
a sufficient certificate for nonzero $\operatorname{col}(\boldsymbol r_i,\boldsymbol\nu_i)$ is
\begin{equation}\label{eq:attenuation_dissipation}
\dot V_i
+
\frac{1}{\eta_i}
\boldsymbol z_i^\top\boldsymbol z_i-
\eta_i
\boldsymbol\nu_i^\top\boldsymbol\nu_i
<0.
\end{equation}

Substituting the channel dynamics~\eqref{channelpre} into \eqref{eq:attenuation_dissipation} gives \begin{equation}\label{eq:attenuation_BRL_quadratic} 
\begin{bmatrix} 
\boldsymbol r_i\\ \boldsymbol\nu_i \end{bmatrix}^{\!\top} 
\boldsymbol M_i^{\rm ch} 
\begin{bmatrix} 
\boldsymbol r_i\\ \boldsymbol\nu_i \end{bmatrix} <0, 
\end{equation} 
where {\footnotesize\begin{equation}\notag \boldsymbol M_i^{\rm ch} := \begin{bmatrix} \operatorname{sym}\!\left( \boldsymbol H\boldsymbol{\hat A}_{cl} \right) +\eta_i^{-1}\boldsymbol C_{K,i}^{\top}\boldsymbol C_{K,i} & \boldsymbol H\boldsymbol B_i +\eta_i^{-1}\boldsymbol C_{K,i}^{\top}\boldsymbol D_i \\ * & -\eta_i\boldsymbol I +\eta_i^{-1}\boldsymbol D_i^{\top}\boldsymbol D_i \end{bmatrix}. \end{equation}} 

Integrating~\eqref{eq:attenuation_dissipation} over $[0,\tau]$ gives $V_i(\boldsymbol r_i(\tau))-V_i(\boldsymbol r_i(0))
+
\frac{1}{\eta_i}
\int_0^{\tau}\boldsymbol z_i^\top\boldsymbol z_i\,\mathrm{d}t-\eta_i
\int_0^{\tau}\boldsymbol\nu_i^\top\boldsymbol\nu_i\,\mathrm{d}t
\leq0$.
Since $\boldsymbol r_i(0)=\boldsymbol 0$ implies
$V_i(\boldsymbol r_i(0))=0$ and
$V_i(\boldsymbol r_i(\tau))\geq0$, the uniform negative-definiteness of $\boldsymbol M_i^{\rm ch}$ gives a strict gain margin as $\tau\rightarrow\infty$.

Since $\boldsymbol M_i^{\rm ch}$ is symmetric, \eqref{eq:attenuation_BRL_quadratic} holds for every nonzero $\operatorname{col}(\boldsymbol r_i,\boldsymbol\nu_i)$ if and only if $\boldsymbol M_i^{\rm ch}\prec\boldsymbol 0$. Since $\boldsymbol M_i^{\rm ch}=\begin{bmatrix} \operatorname{sym}\!\left( \boldsymbol H\boldsymbol{\hat A}_{cl} \right) & \boldsymbol H\boldsymbol B_i \\ * & -\eta_i\boldsymbol I \end{bmatrix}+\eta_i^{-1}\begin{bmatrix} \boldsymbol C_{K,i}^{\top}\\ \boldsymbol D_i^{\top} \end{bmatrix}\begin{bmatrix} \boldsymbol C_{K,i}^{\top}\\ \boldsymbol D_i^{\top} \end{bmatrix}^\top$ and $\eta_i>0$, the Schur complement gives\cite{boyd1994lmi} {\small\begin{equation}\notag 
\boldsymbol M_i^{\rm ch}\prec\boldsymbol 0\Longleftrightarrow \boldsymbol M_i^{\rm ch,Sch}=\begin{bmatrix} \operatorname{sym}\!\left( \boldsymbol H\boldsymbol{\hat A}_{cl} \right) & \boldsymbol H\boldsymbol B_i & \boldsymbol C_{K,i}^{\top} \\ * & -\eta_i\boldsymbol I & \boldsymbol D_i^{\top} \\ * & * & -\eta_i\boldsymbol I \end{bmatrix} \prec\boldsymbol 0. 
\end{equation}}

Because both
$\boldsymbol{\hat A}_{cl}(\boldsymbol K)
=\boldsymbol{\hat A}-\boldsymbol{\hat B}\boldsymbol K$
and $\boldsymbol C_{K,i}$ depend on $\boldsymbol K$,
$\boldsymbol M_i^{\rm ch,Sch}$ is a BMI in
$(\boldsymbol H,\boldsymbol K,\eta_i)$.
A common $\boldsymbol H$ is imposed across all channels and the passivity
constraint as a sufficient, conservative restriction to obtain a single convex state-feedback synthesis after the change of variables
$\boldsymbol X=\boldsymbol H^{-1}$ and
$\boldsymbol Y=\boldsymbol K\boldsymbol X$, and the congruence transformation
$\operatorname{diag}(\boldsymbol X,\boldsymbol I,\boldsymbol I)$ .
\begin{equation}\notag
\boldsymbol\Phi_i
:=
\begin{bmatrix}
\boldsymbol M_{11}
&
\boldsymbol B_i
&
\left(
\boldsymbol C_i\boldsymbol X
-\boldsymbol D_{u,i}\boldsymbol Y
\right)^\top
\\
*
&
-\eta_i\boldsymbol I
&
\boldsymbol D_i^\top
\\
*
&
*
&
-\eta_i\boldsymbol I
\end{bmatrix}
\prec\boldsymbol 0.
\end{equation}

For prescribed passivity margin $\rho_h>0$ and decay rate $\beta>0$, the
attenuation controller is obtained from
{\small
\begin{equation}\label{eq:dynamic_attenuation_synthesis}
\begin{aligned}
\min_{\boldsymbol X,\boldsymbol Y,\{\eta_i\}}
\quad&
\kappa_f\eta_f+\kappa_d\eta_d+\kappa_h\eta_h
+\kappa_G\eta_G+\kappa_T\eta_T
+\mathcal R(\boldsymbol X,\boldsymbol Y)
\\
\mathrm{s.t.}\quad
&
\boldsymbol X\succ\boldsymbol 0,
\qquad
\boldsymbol\Psi^{\rm p}\preceq\boldsymbol 0,
\qquad
\boldsymbol M_{11}\preceq-2\beta\boldsymbol X,
\\
&
\boldsymbol\Phi_i\prec\boldsymbol 0, \qquad\eta_i>0,
\qquad
i\in\{f,d,h,G,T\}.
\end{aligned}
\end{equation}
}

\noindent Here, $\kappa_i>0$ specify the relative channel priorities, while
$\mathcal R$ denotes an SDP-representable convex regularization used for
numerical controller selection. For fixed $\beta$, $\rho_h$, weights and regularization
parameters, \eqref{eq:dynamic_attenuation_synthesis} is a convex semidefinite
program, and
$\boldsymbol K_{\rm att}=\boldsymbol Y\boldsymbol X^{-1}$.

The resulting controller preserves input-strict passivity of the DHS port and
exponential stability of the augmented DHS, while simultaneously attenuating
$\boldsymbol G_f$, $\boldsymbol G_d$, $\boldsymbol G_h$, and the two
DHS-regulation channels $\boldsymbol G_G$ and $\boldsymbol G_T$, with the
direct frequency-response term unchanged.

\section{Numerical Experiments}
\label{sec:numerical}

\begin{figure}[htbp]
    \centering
    \includegraphics[width=\columnwidth]{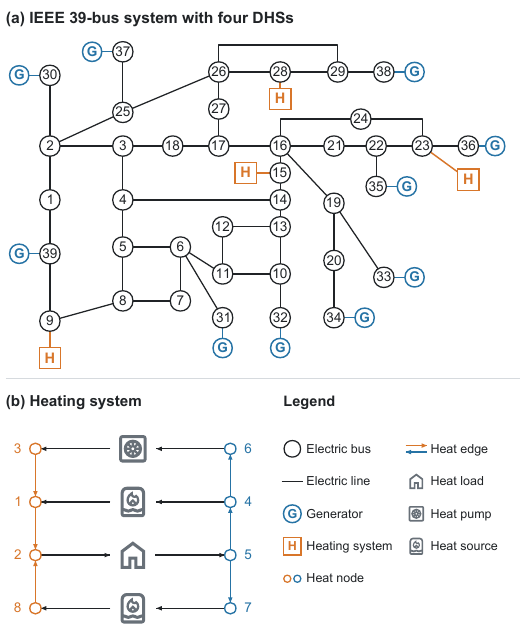}
    \caption{Cross-model validation platform comprising the dynamic modified
    IEEE 39-bus system in \textsc{ANDES} and four FV DHSs connected through
    HPs at buses $9$, $15$, $23$, and $28$.}
    \label{fig:dhstopo}
\end{figure}

High-fidelity validation uses a coupled electro--thermal platform comprising a
modified IEEE 39-bus dynamic model implemented in \textsc{ANDES}, an
open-source platform for electromechanical power-system dynamic
simulation~\cite{cui2020hybrid}, and four finite-volume DHS models retaining
distributed thermal transport, storage, and pipe heat losses~\cite{10380765}.
Following~\cite{qin2025_tpwrs_coupled}, each DHS comprises one HP, two
conventional heat generators, and one controllable load, with the HPs connected
to buses $9$, $15$, $23$, and $28$, as shown in Fig.~\ref{fig:dhstopo}.
Thermal transport paths are spatially
discretized into finite-volume cells, yielding richer DHS
dynamics than the lumped control-oriented model.
Controllers synthesized from \eqref{origindy} are frozen and deployed on the
\textsc{ANDES}--finite-volume platform without retuning, thereby testing
closed-loop performance under electrical and thermal dynamics omitted from the
synthesis model.

$\boldsymbol K_{\rm att}$ is obtained from the multi-channel attenuation
synthesis in \eqref{eq:dynamic_attenuation_synthesis}. The passivity
baseline $\boldsymbol K_{\rm base}$ is a feasible solution of
\eqref{eq:baseline_passivity_synthesis} that satisfies the same passivity and
decay-rate requirements but is not optimized with respect to the attenuation
objectives. Direct frequency-responsive HP control, hereafter denoted as
\emph{Direct HP}, sets $\boldsymbol\gamma^H=\boldsymbol 0$, retaining only the
frequency-responsive component in~\eqref{heatpumpcopcal}. Together, these three
cases separate the DHS-regulation benefit of thermal-state-informed HP control,
the grid-facing behavior of Direct HP, and the additional benefit of
dynamic-port shaping beyond passivity feasibility alone.

\subsection{Verification of the Synthesized Port Properties}
\label{subsec:frozen_frequency_properties}
\begin{figure}[htbp] 
\centering 
\includegraphics[width=\columnwidth]{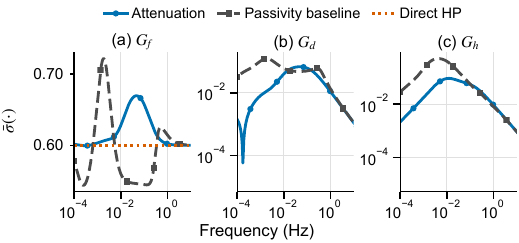}
\caption{Maximum singular values of the grid-facing channels
$\boldsymbol G_f$, $\boldsymbol G_d$, and $\boldsymbol G_h$ under the frozen
controllers synthesized on the control-oriented model.}
\label{fig:frozen_port_channels} 
\end{figure}

\begin{figure}[htbp]
    \centering
    \includegraphics[width=\columnwidth]
    {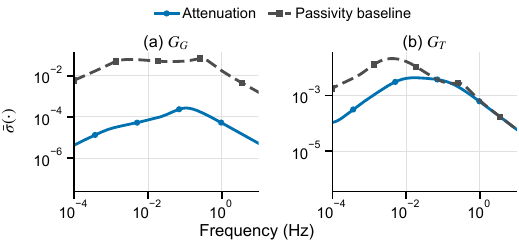}
    \caption{Maximum singular values of the DHS-regulation channels
    $\boldsymbol G_G$ and $\boldsymbol G_T$
    under the frozen controllers synthesized on the control-oriented model.}
    \label{fig:frozen_error_channels}
\end{figure}

We first verify the port properties enforced by the synthesis on the
control-oriented DHS model. Fig.~\ref{fig:frozen_port_channels} shows the
grid-facing channels $\boldsymbol G_f$, $\boldsymbol G_d$, and
$\boldsymbol G_h$ in~\eqref{threeportsfdh}. The passivity baseline exhibits a
pronounced DHS-state-dependent contribution around the dominant DHS time
scale, approximately in the $10^{-2}$-Hz range, overlapping with the time
scale of secondary frequency control. Passivity alone does not
constrain the phase of this contribution; over part of the frequency range,
$\boldsymbol G_d$ can therefore oppose the frequency-responsive component and
partially offset the damping associated with $\boldsymbol\gamma^E$. The
attenuation controller reduces the gain of $\boldsymbol G_d$, keeping
the full grid-facing map
$\boldsymbol G_f=\boldsymbol\gamma^E+\boldsymbol G_d$ close to the response of
direct frequency-responsive HP control. It also suppresses
$\boldsymbol G_h$, thereby limiting heat-disturbance propagation to the
electrical port. Together, these channels quantify the DHS-state-dependent
electrical dynamics and heat-disturbance propagation that overlap with
secondary frequency control. Fig.~\ref{fig:frozen_error_channels} further shows
that the DHS-regulation channels $\boldsymbol G_G$ and
$\boldsymbol G_T$ are reduced relative to the passivity baseline.

\vspace{-0.3cm}

The attenuation design minimizes a weighted sum of certified gain bounds over
the full frequency range. Consistent with the high-frequency behavior established
in Section~\ref{sec:loop_shaping}, $\boldsymbol G_f(j\omega)$ approaches
$\boldsymbol\gamma^E$ at high frequencies, so the response associated
with primary frequency control is largely preserved. The shaping is therefore
most pronounced at the slower DHS time scales overlapping with secondary
frequency control.

\vspace{-0.3cm}

\subsection{Multi-Channel Attenuation and DHS Regulation}
\label{subsec:attenuation_benefit}

We next assess whether the synthesized attenuation of
$\boldsymbol G_h$, $\boldsymbol G_G$, and $\boldsymbol G_T$ carries over to
the coupled electromechanical EPS--finite-volume DHS platform. No external EPS
disturbance is applied. Each DHS is subjected to the sinusoidal heat-demand
disturbance $q^L=10.4\sin(2\pi f_ht)\,\mathrm{MW}_{\rm th}$, smoothly
introduced over $60$--$120$~s, for
$f_h\in\{0.016,0.03,0.05,0.10\}$~Hz. All cases retain nominal inertia,
native TGOV1N primary control, common AGC, and fixed DHS mass flow. The
simulation horizon is $t_f=\max\{600,120+4/f_h\}$~s, and the final three
disturbance cycles define the evaluation window $\mathcal T_w$.

To isolate the oscillatory regulation response, the mean of the
error signal is removed over $\mathcal T_w$, whose duration is $T_w$.
For controller $j$ and DHS $a$,
$\widetilde e_{\ell,a}^{\,j}(t):=e_{\ell,a}^{\,j}(t)
-T_w^{-1}\int_{\mathcal T_w}e_{\ell,a}^{\,j}(t')\,{\rm d}t'$,
$\ell\in\{T,G1,G2\}$.
The temperature- and generation-regulation metrics are
$R_T^j:=\bigl[(4T_w)^{-1}\sum_{a=1}^{4}\int_{\mathcal T_w}
\|\widetilde e_{T,a}^{\,j}(t)\|_2^2\,{\rm d}t\bigr]^{1/2}$ and
$R_G^j:=\bigl[(4T_w)^{-1}\sum_{a=1}^{4}\int_{\mathcal T_w}
\|[\widetilde e_{G1,a}^{\,j}(t),\widetilde e_{G2,a}^{\,j}(t)]^\top\|_2^2
\,{\rm d}t\bigr]^{1/2}$. Smaller values correspond to smaller
DHS-regulation errors.

\subsubsection{DHS-Regulation Channels}
\label{subsubsec:temperature_attenuation}

Table~\ref{tab:dhs_regulation_attenuation} evaluates the regulation channels
identified by $\boldsymbol G_G$ and $\boldsymbol G_T$ in
Fig.~\ref{fig:frozen_error_channels}. The attenuation controller gives the
smallest $R_T$ and $R_G$ over all disturbance frequencies. Relative
to direct frequency-responsive HP control, $\boldsymbol K_{\rm att}$ reduces
the temperature-regulation metric by $17$--$28\%$ and the generation-sharing
metric by $4$--$27\%$. Thus, thermal-state-informed HP control allows HP
modulation to contribute directly to DHS temperature regulation and
heat-generation sharing, rather than leaving the resulting thermal imbalance
entirely to the conventional DHS regulation layer. The much larger errors
obtained with $\boldsymbol K_{\rm base}$ further show that passivity and
closed-loop stability alone do not guarantee effective DHS regulation under
persistent heat-demand disturbances. The agreement with the frequency-domain
ordering of $\boldsymbol G_G$ and $\boldsymbol G_T$ in Fig.~\ref{fig:frozen_error_channels} provides
evidence that the multi-channel attenuation objective translates to improved
DHS regulation on the finite-volume models.

\begin{table}[htbp]
\centering
\caption{FV-DHS regulation metrics under sinusoidal heat-demand disturbances.}
\label{tab:dhs_regulation_attenuation}
\scriptsize
\setlength{\tabcolsep}{4.0pt}
\begin{tabular}{c|c|ccc}
\hline
$f_h$ (Hz) & Metric
& $\boldsymbol K_{\rm att}$
& Direct HP
& $\boldsymbol K_{\rm base}$\\
\hline
0.016
& $R_T$ (K)
& $\mathbf{0.3377}$ & $0.4086$ & $1.2678$\\
& $R_G$
& $\mathbf{7.50{\times}10^{-4}}$
& $7.83{\times}10^{-4}$
& $0.2162$\\
\hline
0.030
& $R_T$ (K)
& $\mathbf{0.2825}$ & $0.3802$ & $0.6641$\\
& $R_G$
& $\mathbf{8.02{\times}10^{-4}}$
& $9.63{\times}10^{-4}$
& $0.1167$\\
\hline
0.050
& $R_T$ (K)
& $\mathbf{0.2097}$ & $0.2911$ & $0.3369$\\
& $R_G$
& $\mathbf{8.04{\times}10^{-4}}$
& $1.07{\times}10^{-3}$
& $0.06310$\\
\hline
0.100
& $R_T$ (K)
& $\mathbf{0.1644}$ & $0.2106$ & $0.1656$\\
& $R_G$
& $\mathbf{9.88{\times}10^{-4}}$
& $1.35{\times}10^{-3}$
& $0.03873$\\
\hline
\end{tabular}
\end{table}

\begin{figure}[htbp]
\centering
\includegraphics[width=\columnwidth]{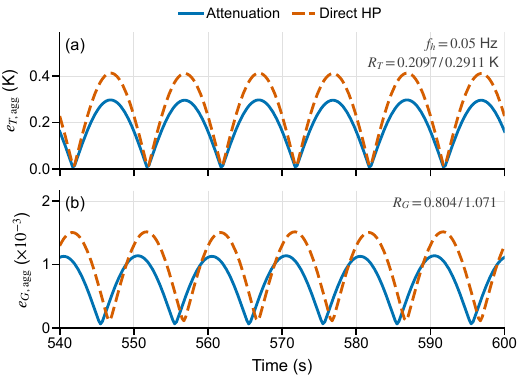}
\caption{Representative DHS-regulation response at $f_h=0.05$~Hz.
\rev{Each curve is the instantaneous root-mean-square, across the four DHSs, of the
mean-removed temperature-regulation and generation-sharing errors.}
Lower overall oscillation levels agree with Table~\ref{tab:dhs_regulation_attenuation}.}
\label{fig:dhs_regulation_transient}
\end{figure}

To illustrate the corresponding time-domain behavior,
Fig.~\ref{fig:dhs_regulation_transient} shows a representative case at
$f_h=0.05$~Hz. The attenuation controller yields smaller aggregate
temperature-regulation oscillations and a lower overall generation-regulation
oscillation level than direct frequency-responsive HP control, consistent with
the reductions in $R_T$ and $R_G$ in
Table~\ref{tab:dhs_regulation_attenuation}.

\subsubsection{Heat-Disturbance Propagation to the EPS}
\label{subsubsec:disturbance_leakage}

The same experiment isolates the heat-disturbance pathway
$\boldsymbol G_h$ in~\eqref{portmapheat}. For controller $j$, define the
aggregate HP electrical-power variation induced by DHS-state feedback as
\[
\Delta p_{d,\Sigma}^{H,j}(t)
:=
\sum_{a=1}^{4}
\gamma_a^H
\left[
S_a^{H,j}(t)-S_a^{H,j}(0)
\right].
\]

For $\mathcal T_w$ with duration $T_w$, let
$\widetilde{\Delta p}_{d,\Sigma}^{H,j}(t):=\Delta p_{d,\Sigma}^{H,j}(t)
-T_w^{-1}\int_{\mathcal T_w}\Delta p_{d,\Sigma}^{H,j}(t')\,{\rm d}t'$
and
$\widetilde{\Delta f}^{\,j}(t):=\Delta f^{\,j}(t)
-T_w^{-1}\int_{\mathcal T_w}\Delta f^{\,j}(t')\,{\rm d}t'$.
The DHS-to-EPS propagation and resulting frequency oscillation are measured by
$R_{P,h}^{j}:=\bigl[T_w^{-1}\int_{\mathcal T_w}
|\widetilde{\Delta p}_{d,\Sigma}^{H,j}(t)|^2\,{\rm d}t\bigr]^{1/2}$ and
$R_{f,h}^{j}:=\bigl[T_w^{-1}\int_{\mathcal T_w}
|\widetilde{\Delta f}^{\,j}(t)|^2\,{\rm d}t\bigr]^{1/2}$, reported in MW and mHz,
respectively. For direct frequency-responsive HP control,
$\boldsymbol\gamma^H=\boldsymbol 0$, and hence
$\Delta p_{d,\Sigma}^{H,j}\equiv0$ by construction. This provides structural
isolation of heat-side disturbances from the electrical port, whereas the
comparison below evaluates how effectively the two thermal-state-informed
controllers contain this pathway while retaining DHS regulation.
\begin{table}[htbp]
\centering
\caption{HEAT-DISTURBANCE PROPAGATION UNDER THERMAL-STATE-INFORMED HP CONTROL. Each
entry is $R_{P,h}$ (MW)/$R_{f,h}$ (mHz).}
\label{tab:heat_disturbance_leakage}
\scriptsize
\setlength{\tabcolsep}{7pt}
\begin{tabular}{c|cc}
\hline
$f_h$ (Hz)
& $\boldsymbol K_{\rm att}$
& $\boldsymbol K_{\rm base}$\\
\hline
0.016
& $\mathbf{32.423/8.975}$
& $121.713/33.699$\\
\hline
0.030
& $\mathbf{27.124/7.951}$
& $63.750/18.684$\\
\hline
0.050
& $\mathbf{20.131/6.593}$
& $32.341/10.591$\\
\hline
0.100
& $\mathbf{15.784/6.299}$
& $15.894/6.320$\\
\hline
\end{tabular}
\end{table}

Table~\ref{tab:heat_disturbance_leakage} shows that
$\boldsymbol K_{\rm att}$ reduces heat-side disturbance propagation to the
electrical interface relative to the passivity baseline at every reported
disturbance frequency. The reduction is strongest where the baseline exhibits
large DHS dynamic gains: $R_{P,h}$ decreases from $121.713$ to
$32.423$~MW at $0.016$~Hz and from $63.750$ to $27.124$~MW at
$0.030$~Hz, corresponding to reductions of $73.4\%$ and $57.5\%$,
respectively. The associated EPS frequency oscillations decrease from
$33.699$ to $8.975$~mHz and from $18.684$ to $7.951$~mHz. At
$0.10$~Hz, the two controllers become nearly indistinguishable. These results associate
the attenuation of $\boldsymbol G_h$ with reduced heat-side disturbance
propagation at the electrical interface, supporting the clean grid-facing
response objective.

\subsection{Cross-Model Full-Period Validation}
\label{subsec:full_period_deployment}

We finally assess the complete multi-timescale closed loop on the coupled
electromechanical IEEE 39-bus--finite-volume DHS platform. At $t=60$~s, the EPS is subjected to
a persistent $600$-MW load increase, while each DHS receives a
$10.4$-$\mathrm{MW}_{\rm th}$ heat-demand increase ramped over
$60$--$120$~s. Native primary frequency control remains active throughout,
and AGC is enabled at $t=120$~s. All cases use nominal inertia, fixed DHS mass
flow, and a $5000$-s simulation horizon. The closed-loop objectives are
$\Delta f\rightarrow0$, $\boldsymbol e_G\rightarrow\boldsymbol0$, and
$e_T\rightarrow0$.

To quantify the aggregate DHS regulation burden without mean removal, define
\[
r_e^j(t)
:=
\left[
\frac{1}{4}
\sum_{a=1}^{4}
\left(
\|\boldsymbol e_{G,a}^j(t)\|_2^2
+
\bigl(e_{T,a}^j(t)\bigr)^2
\right)
\right]^{1/2},
\]
where $\boldsymbol e_{G,a}^j$ collects the two generation-sharing errors in
DHS $a$. The full-period regulation burden and peak error over $60\leq t\leq5000$~s are
\[
J_{\rm DHS}^j
:=
\left[
\int_{60}^{5000}
\bigl(r_e^j(t)\bigr)^2\,{\rm d}t
\right]^{1/2},
\qquad
r_{e,\max}^j
:=
\max_t r_e^j(t).
\]
The frequency-recovery time $t_f^{10^{-4}}$ is the elapsed time from $t=60$~s to the first instant after which
$|\Delta f|\leq10^{-4}$~Hz continuously for $300$~s.

\begin{table}[t]
\centering
\caption{Full-period closed-loop performance under simultaneous electrical
and thermal disturbances.}
\label{tab:full_period_deployment}
\scriptsize
\setlength{\tabcolsep}{3.8pt}
\begin{tabular}{c|cccc}
\hline
Controller
& $\Delta f_{\rm nadir}$ (Hz)
& $t_f^{10^{-4}}$ (s)
& $J_{\rm DHS}$
& $r_{e,\max}$
\\
\hline
$\boldsymbol K_{\rm att}$
& $\mathbf{-0.215303}$
& $\mathbf{1162.75}$
& $\mathbf{1.43560}$
& $\mathbf{0.23406}$
\\
Direct HP control
& $-0.216062$
& $1172.75$
& $1.44767$
& $0.24232$
\\
$\boldsymbol K_{\rm base}$
& $-0.216732$
& $2607.00$
& $13.94711$
& $1.74154$
\\
\hline
\end{tabular}
\end{table}

All three configurations remain bounded and converge to the required
electrical and thermal equilibrium, providing empirical cross-model support for the
stability properties established in
Lemma~\ref{lem:dhs_cl_properties}--\ref{lem:nominal_coupled_equilibrium} and
Theorem~\ref{thm:nominal_interconnection_stability}. As shown in
Table~\ref{tab:full_period_deployment}, the attenuation controller remains
close to direct frequency-responsive HP control in grid-frequency response:
the nadir differs by only $0.759$~mHz, while the frequency-recovery times
differ by $10$~s. This close agreement is the desired outcome: shaping
preserves the electrical behavior of Direct HP during frequency restoration. At the same time, $\boldsymbol K_{\rm att}$ reduces
the full-period DHS burden and peak regulation error.

The contrast with the passivity baseline is larger.
$\boldsymbol K_{\rm att}$ reduces $J_{\rm DHS}$ by $89.7\%$ and
$r_{e,\max}$ by $86.6\%$ relative to $\boldsymbol K_{\rm base}$, while
shortening the frequency-recovery time from $2607.00$ to $1162.75$~s.
This recovery-time comparison concerns frequency restoration with EPS AGC.
Passivity feasibility under the theorem assumptions is sufficient for stability
but does not by itself ensure multi-timescale regulation performance. 

\section{Conclusion}
\label{sec:conclusion}

This paper formulated the DHS under thermal-state-informed HP control
as a grid-facing dynamic port and developed a mixed
$\mathcal H_\infty$--passivity framework for DHS-regulator synthesis.
The stability analysis spans primary frequency control, secondary frequency
control, and DHS regulation. Multi-channel attenuation limits
DHS-state-dependent electrical dynamics, heat-disturbance propagation, and
DHS-regulation errors on the time scale of secondary frequency control.
Cross-model validation on the IEEE 39-bus system with four finite-volume DHS
models shows improved DHS regulation while preserving the grid-facing response
of direct frequency-responsive HP control during secondary frequency control.
\appendices

\section*{Proof of Lemma~\ref{lem:eps_frequency_restoration}}

At equilibrium, \eqref{eq:agc_state} gives
$\boldsymbol\alpha^\top\boldsymbol\omega^\star=0$.
By Assumption~\ref{ass:eps_equilibrium}, the steady-state bus frequencies are
synchronized, i.e., $\boldsymbol\omega^\star
=
\omega^{com}\boldsymbol 1_{n_{\mathcal E}}$.
Since
$\boldsymbol\alpha^\top\boldsymbol 1_{n_{\mathcal E}}=1$, it follows that
$\omega^{com}=0$. Hence,
$\boldsymbol\omega^\star=\boldsymbol 0$.
\hfill $\square$

\section*{Proof of Lemma~\ref{lem:eps_passive}}

Let $\boldsymbol S_{HP}$ denote the HP-bus selection matrix, such that
$\boldsymbol y_1
=
\boldsymbol\omega^{HP}
=
\boldsymbol S_{HP}^\top\boldsymbol\omega$,
and define
$\tilde{\boldsymbol P}^G
:=
\boldsymbol P^G-\boldsymbol P^{G\star}$.

The passivity relation below is written in equilibrium-referenced deviations, with $\tilde{\boldsymbol u}_1=-(\boldsymbol p^H-\boldsymbol p^{H\star})$. Consider the equilibrium-referenced network~\eqref{busfre1}-\eqref{busfre2} storage of
$V_{\mathrm{net}}
=
\frac{1}{2}\boldsymbol\omega^\top\boldsymbol M\boldsymbol\omega
+
U(\boldsymbol x_\theta)-U(\boldsymbol x_\theta^\star)
-
\nabla U(\boldsymbol x_\theta^\star)^\top
(\boldsymbol x_\theta-\boldsymbol x_\theta^\star)$,
where $U$ is the network potential energy. Under the angle condition in
Assumption~\ref{ass:eps_equilibrium}, $V_{\mathrm{net}}$ is locally
nonnegative. Using the swing equations, the equilibrium power-balance
condition, and the fact that $\boldsymbol P^L$ is constant gives
$\dot V_{\mathrm{net}}
=
\boldsymbol y_1^\top\tilde{\boldsymbol u}_1
+
\boldsymbol\omega^\top\tilde{\boldsymbol P}^G
-
\boldsymbol\omega^\top\boldsymbol D\boldsymbol\omega$.
By Assumption~\ref{ass:net_supply_passivity}, the
generation--governor--AGC subsystem admits an equilibrium-referenced storage
function $V_g$ satisfying that
$\dot V_g
\le
-\boldsymbol\omega^\top\tilde{\boldsymbol P}^G
-\rho_g\|\boldsymbol\omega\|^2$.
Defining
$V_{\mathrm e}:=V_{\mathrm{net}}+V_g$ for~\eqref{gen_dyn}-\eqref{eq:primary_secondary_control}, yields
$\dot V_{\mathrm e}
\le
\boldsymbol y_1^\top\tilde{\boldsymbol u}_1
-
\boldsymbol\omega^\top\boldsymbol D\boldsymbol\omega
-
\rho_g\|\boldsymbol\omega\|^2$.
Since $\boldsymbol y_1=\boldsymbol S_{HP}^\top\boldsymbol\omega$ and
$\|\boldsymbol y_1\|\le\|\boldsymbol S_{HP}\|_2\|\boldsymbol\omega\|$, it follows that
$\dot V_{\mathrm e}
\le
\boldsymbol y_1^\top\tilde{\boldsymbol u}_1
-
\frac{\rho_g}{\|\boldsymbol S_{HP}\|_2^2}\|\boldsymbol y_1\|^2$.
Therefore, the EPS is locally output strictly passive from
$\tilde{\boldsymbol u}_1$ to $\boldsymbol y_1$, with $\rho_e=\rho_g/\|\boldsymbol S_{HP}\|_2^2$.
\hfill $\square$

\section*{Proof of Theorem~\ref{thm:dhs_opt_nominal}}
For \textbf{OPT-1} in \eqref{opt1_nominal_all}, the KKT conditions are
$\boldsymbol W^G\boldsymbol q^G+\boldsymbol B_G^\top\boldsymbol\mu=\boldsymbol 0$,
$\boldsymbol A^\top\boldsymbol\mu=\boldsymbol 0$,
together with primal feasibility \eqref{opt1_nominal_con}. Since the DHS
Kirchhoff matrix satisfies
$\ker(\boldsymbol A^\top)=\mathrm{span}\{\boldsymbol V\boldsymbol 1\}$ and $\boldsymbol B_G^\top\boldsymbol V\boldsymbol 1=(\rho C_p)^{-1}\boldsymbol 1$, all conventional
marginal costs are equal, which is equivalent to
$\boldsymbol W_G^M\boldsymbol q^G=\boldsymbol 0$.
Strict convexity of \eqref{opt1_nominal_obj} in $\boldsymbol q^G$ gives
uniqueness. For \textbf{OPT-2} in \eqref{opt2_nominal_all}, every feasible
profile can be written as
$\boldsymbol T(\alpha)=\boldsymbol A^\dagger(\cdot)+\alpha\boldsymbol 1$.
Differentiating the scalar quadratic objective in \eqref{opt2_nominal_obj} with
respect to $\alpha$ gives
$\boldsymbol 1^\top\boldsymbol W^T\boldsymbol T^\star=0$, and strict convexity
in the scalar direction $\alpha$ gives uniqueness. \hfill $\square$

\section*{Proof of Lemma~\ref{lem:dhs_cl_properties}}
For constant inputs, \eqref{closeloopdhs_nominal} is the affine LTI system
$\dot{\boldsymbol x}=\boldsymbol{\hat A}_{cl}(\boldsymbol K)\boldsymbol x+\boldsymbol b$.
Since $\boldsymbol{\hat A}_{cl}(\boldsymbol K)$ is exponentially stable, it is
nonsingular and the equilibrium is unique and globally attractive, which gives
\eqref{eq:dhs_equilibrium_xstar}. At equilibrium,
$\dot{\boldsymbol\xi}=\boldsymbol e$ implies $\boldsymbol e^\star=\boldsymbol 0$.
Therefore the last component of \eqref{eq:error_nominal} yields
$\boldsymbol 1_{n_T}^\top\boldsymbol W^T\boldsymbol T^\star=0$, which is exactly
$S^{H\star}=0$ in \eqref{eq:coord_signal_nominal}; substituting this into
\eqref{eq:y2_nominal} gives \eqref{eq:C0_xstar_zero}. Since
\eqref{eq:C0_xstar_zero} holds for arbitrary constant
$\bar{\boldsymbol\omega}^{HP}$ and $\bar{\boldsymbol w}^h$, the identities in
\eqref{eq:C0_Acl_inverse_zero} follow from \eqref{eq:dhs_equilibrium_xstar}.
Subtracting the equilibrium from \eqref{closeloopdhs_nominal} gives
\eqref{eq:dev_dyn_shifted_A}. If $\tilde{\boldsymbol u}_2\equiv\boldsymbol 0$,
any zero-input equilibrium satisfies
$\boldsymbol{\hat A}_{cl}(\boldsymbol K)\tilde{\boldsymbol x}=\boldsymbol 0$,
and hence $\tilde{\boldsymbol x}=\boldsymbol 0$. \hfill $\square$

\section*{Proof of Lemma~\ref{lem:nominal_coupled_equilibrium}}
Lemma~\ref{lem:eps_frequency_restoration} gives
$\boldsymbol\omega^\star=\boldsymbol 0$ for any EPS equilibrium under AGC.
Lemma~\ref{lem:dhs_cl_properties} gives a unique DHS equilibrium with
$\boldsymbol e^\star=\boldsymbol 0$. Therefore,
$\boldsymbol W_G^M\boldsymbol q^{G\star}=\boldsymbol 0$ and
$\boldsymbol 1_{n_T}^\top\boldsymbol W^T\boldsymbol T^\star=0$. In~\eqref{firstorder}, $\boldsymbol\omega^\star=\boldsymbol 0$ and
$S^{H\star}=0$ imply $\boldsymbol p^{H\star}=\boldsymbol 0$ and hence
$\boldsymbol q^{P\star}=\boldsymbol 0$. Theorem~\ref{thm:dhs_opt_nominal}
then gives the stated DHS optimality, while Lemma~\ref{lem:eps_frequency_restoration}
gives the stated EPS frequency-restoration property. \hfill $\square$

\section*{Proof of Theorem~\ref{thm:nominal_interconnection_stability}}
By Lemma~\ref{lem:eps_passive}, where $\boldsymbol p^{H\star}=\boldsymbol 0$ by
Lemma~\ref{lem:nominal_coupled_equilibrium} and hence
$\tilde{\boldsymbol u}_1={\boldsymbol u}_1$,
$\dot V_e\le {\boldsymbol y}_1^\top{\boldsymbol u}_1-
\rho_e\|{\boldsymbol y}_1\|^2$.
By the input-strict passivity of the DHS port,
$\dot V_h\le {\boldsymbol y}_2^\top{\boldsymbol u}_2-
\rho_h\|{\boldsymbol u}_2\|^2$.
Using the interconnection
${\boldsymbol u}_2={\boldsymbol y}_1$ and
${\boldsymbol u}_1=-{\boldsymbol y}_2$ gives
$\dot V_e+
\dot V_h\le-\rho_g\|\boldsymbol\omega\|^2-\rho_h\|\boldsymbol\omega^{HP}\|^2\le0$,
using the stronger bound in the proof of Lemma~\ref{lem:eps_passive}.
In a sufficiently small compact invariant storage sublevel set, zero dissipation implies $\boldsymbol\omega=0$. Hence angles and $g$ are constant. Since $\boldsymbol{\hat A}_{cl}$ is Hurwitz, every complete bounded zero-input DHS trajectory is its equilibrium. The swing and governor equations then fix generation and $g$; the connected network's angle condition gives local uniqueness of the relative angles.
LaSalle's invariance principle, together with the zero-input property in
Lemma~\ref{lem:dhs_cl_properties}, yields local asymptotic stability of the
equilibrium in Lemma~\ref{lem:nominal_coupled_equilibrium}. \hfill $\square$

\section*{Congruence Transformations of the Passivity and Decay-Rate Conditions}
Apply the congruence transformation
$\operatorname{diag}(\boldsymbol X,\boldsymbol I)$ to
\eqref{eq:KYP_BMI_shifted_A}. Since
$\boldsymbol{\hat A}_{cl}(\boldsymbol K)
=\boldsymbol{\hat A}-\boldsymbol{\hat B}\boldsymbol K$, we have
$\boldsymbol X
\bigl(
\boldsymbol{\hat A}_{cl}^{\top}\boldsymbol H
+\boldsymbol H\boldsymbol{\hat A}_{cl}
\bigr)
\boldsymbol X
=
\operatorname{sym}
\bigl(
\boldsymbol{\hat A}\boldsymbol X
-\boldsymbol{\hat B}\boldsymbol Y
\bigr)
=:\boldsymbol M_{11}$,
and
$\boldsymbol X
\bigl(
\boldsymbol H\boldsymbol{\hat B}_{\omega}
-\boldsymbol C_0^\top
\bigr)
=
\boldsymbol{\hat B}_{\omega}
-\boldsymbol X\boldsymbol C_0^\top
=:\boldsymbol M_{12}$.
Hence, \eqref{eq:KYP_BMI_shifted_A} is equivalent to
\eqref{eq:LMI_shifted}. Similarly, applying the same transformation to
\eqref{eq:decay_rate_BMI} gives
$\boldsymbol M_{11}+2\beta\boldsymbol X\preceq0$,
which is \eqref{eq:decay_rate_LMI_short}. The controller is recovered as
$\boldsymbol K=\boldsymbol Y\boldsymbol X^{-1}$.

\bibliographystyle{IEEEtran}
\bibliography{CDC-port1}

\end{document}